\documentclass[pre,10pt]{revtex4}

\usepackage{amsmath}    
\usepackage{graphicx}   
\usepackage{verbatim}   
\usepackage{color}      
\usepackage{subfigure}  
\usepackage{hyperref}   
\usepackage{pstricks}
\usepackage{epstopdf}

\newcommand{\be}{\begin{equation}}
\newcommand{\ee}{\end{equation}}
\newcommand{\bea}{\begin{eqnarray}}
\newcommand{\eea}{\end{eqnarray}}
\newcommand{\bs}{\boldsymbol}
\newcommand{\dif}{{\rm d}}
\newcommand{\ddif}[2]{\frac{{\rm d}#1}{{\rm d}#2}}
\newcommand{\exx}[1]{{\rm e}^{#1}}
\newcommand{\ind}[1]{_{\text{ #1}}}
\newcommand{\Eq}[1]{Eq.~(\ref{#1})}
\newcommand{\ii}{{\rm i}}
\newcommand{\Hop}{{\rm H}}
\newcommand{\op}[1]{{\rm #1}}
\newcommand{\nn}{\nonumber}
\newcommand{\aop}{\op{a}^{ }}
\newcommand{\aopd}{\op{a}^\dagger}
\newcommand{\rf}[1]{Eq.~(\ref{#1})}
\def\e{\textrm{e}}

\def\bs{\boldsymbol}

\begin{document}

\title{
Electrical conductivity of charged particle systems and the Zubarev NSO
method
}

\author{G. R\"opke}
\affiliation{Universit\"at Rostock, Institut f\"ur Physik, 18051 Rostock, Germany}

\date{\today}

\begin{abstract}
One of the fundamental problems in physics which are not rigorously solved yet 
is the statistical mechanics of nonequilibrium processes.
An important contribution to describe irreversible behavior starting from reversible Hamiltonian dynamics 
was given by D. N. Zubarev 
who invented the method of the nonequilibrium statistical operator (NSO). 
We discuss this approach, in particular the extended von Neumann equation, 
and consider as example the electrical conductivity of a charged particle system. 
The selection of the set of relevant observables is considered.
The relation between kinetic theory and linear response theory is shown. 
Using thermodynamic Green functions, a systematic treatment of correlation functions is given, 
but convergence has to be investigated.
Different expressions for the conductivity are compared, and open questions are identified.

\end{abstract}

\maketitle

\section{The Zubarev NSO method}

After the laws of thermodynamics have been formulated  in the 19th century, in particular the definition of entropy 
for systems in thermodynamic equilibrium and the increase of intrinsic entropy in nonequilibrium processes, 
the microscopic approach to the nonequilibrium evolution was first given by Ludwig Boltzmann 
who formulated the kinetic theory of gases \cite{Boltzmann} using the famous Sto{\ss}zahlansatz. 
The question how irreversible evolution in time can be 
obtained from reversible microscopic equations  has been arisen  immediately and controversially discussed.
The rigorous derivation of the kinetic equations from a microscopic description of a system was given only long time afterwards by Bogoliubov \cite{Bogoliubov}
introducing a new additional theorem, the principle of weakening of initial correlations.

A generalization has been given by Zubarev \cite{Zubarev} who invented the method of the nonequilibrium statistical operator (NSO).
This approach has been applied to various problems in nonequilibrium statistical physics, see \cite{ZMR1,ZMR2} and may be considered as 
a unified, fundamental approach to non equilibrium systems which includes different theories such as kinetic theory (KT), linear response theory (LRT), and quantum master equations (QMA). 
We present here the LRT with special application to the electrical conductivity of charged particle systems.
Different expressions are discussed and their relations are given. 
The Ziman, Spitzer, and Kubo-Greenwood expressions are considered.
A Green function approach \cite{ZubarevGF} to evaluate correlation functions is investigated.
Hopping conductivity, convergence, and virial expansions are discussed, and the problem of entropy production is outlined.

An exhaustive review of the Zubarev NSO method and its manifold applications cannot be given here. 
We discuss only a very special application, the evaluation of the electrical conductivity of charge particle systems. For more references see also the recent publications \cite{Luzzi,Kuzemsky,Ryazanov}.

Within statistical mechanics, the thermodynamic state of an ensemble of many-particle systems at time $t$ is described 
by the statistical operator $\rho(t)$.
We assume that the time evolution of the quantum state of the system is given by the Hamiltonian ${\rm H}^t$ which may contain time-dependent external fields.
The von Neumann equation follows as equation of motion for the statistical operator,
\begin{equation}
\label{1.5}
 \frac{\partial}{\partial t}\rho(t) + \frac{\ii}{\hbar} \left[{\rm H}^t,\rho(t) \right]=0.
\end{equation}

The von Neumann equation describes reversible dynamics. The equation of motion is based on the Schr\"odinger equation. Time inversion and conjugate complex means that the first term on the left hand side as well as the second one change the sign, since $\ii \to -\ii$ and both the Hamiltonian and the statistical operator are Hermitean.
However, the von Neumann equation is not sufficient to determine  $\rho(t)$ because it is a first order differential equation,
and an initial value  $\rho(t_0)$ at time $t_0$ is necessary to specify a solution. This problem emerges clearly in equilibrium.\\

{\it Thermodynamic equilibrium}.
By definition, in thermodynamic equilibrium, the thermodynamic state of the system is not changing with time. 
Both, ${\rm H}^t$ and  $\rho(t)$, are not depending on $t$ so that
 \begin{equation}
\frac{\partial}{\partial t}\rho_{\text{eq}}(t)=0.
\end{equation}
The solution of the von Neumann equation in thermodynamic equilibrium becomes trivial,
\begin{equation}
\label{1.5a}
 \frac{\ii}{\hbar} \left[{\rm H},\rho_{\text{eq}} \right]=0.
\end{equation}
The time-independent statistical operator  $ \rho_{\text{eq}}$ commutes with the Hamiltonian. 
We conclude that $ \rho_{\text{eq}}$ depends only on constants of motion ${\rm C}_n$ that commute with ${\rm H}$. But the  von Neumann equation is not sufficient to determine how  $ \rho_{\text{eq}}$ depends on constants of motion ${\rm C}_n$. We need a new additional principle, not included in the Hamiltonian dynamics.

Equilibrium statistical mechanics is based of the following principle to determine 
the statistical operator $\rho_{\text{eq}}$:
Consider the functional (information entropy)
 \begin{equation} 
\label{Sinf}
S_{\text{inf}} [\rho] = - {\rm Tr} \{\rho \ln \rho\} \,
\end{equation} 
for arbitrary  $\rho$ that are consistent with the given conditions
${\rm Tr} \{\rho\} =1$
(normalization) and
\begin{equation}
\label{sceq}
 {\rm Tr} \{\rho\, {\rm C}_n\} = \langle {\rm C}_n \rangle
\end{equation}
(self-consistency conditions).
With this conditions, we vary $\rho$ and determine the maximum of the information entropy for the optimal distribution  $\rho_{\text{eq}}$ so that $\delta S_{\text{inf}} [\rho_{\text{eq}}]=0$. 
As well known, the method of Lagrange multipliers can be used to account for the self-consistency conditions (\ref{sceq}).
The corresponding maximum value for $S_{\text{inf}}[\rho] $
\begin{equation} 
\label{Seq}
S_{\text{eq}} [\rho_{\text{eq}}] = -k_{\text{B}} {\rm Tr} \{\rho_{\text{eq}} \ln \rho_{\text{eq}}\} \,
\end{equation} 
is the equilibrium entropy of the system at given constraints $\langle{\rm C}_n \rangle$,
$k_{\text{B}}
$ is the Boltzmann constant.
The solution of this variational principle leads to the Gibbs ensembles for thermodynamic equilibrium.

As an example, we consider an open system which is in thermal contact and particle exchange with reservoirs. 
The sought-after equilibrium statistical operator has to  obey the given constraints:  
normalization ${\rm Tr} \{\rho\} =1 $, 
 thermal contact with the bath so that $ {\rm Tr} \{\rho \,{\rm H}\} = U$ (internal energy), 
 particle exchange with a reservoir so that for the particle number operator ${\rm N}_c$ of species $c$, the average is given by $ {\rm Tr} \{\rho\, {\rm N}_c\} = n_c \Omega$, where $\Omega$ denotes the volume of the system (we don't use $V$ to avoid confusion with the potential), and $n_c$ the particle density of species $c$.
Looking for the maximum of the information entropy functional
with these constraints, one obtains the grand canonical distribution
\begin{equation}
\label{gr.can}
\rho_{\rm eq}= \frac{\e^{-\beta ({\rm H}-\sum_c \mu_c{\rm N}_c)}}{{\rm Tr}\, \e^{-\beta ({\rm H}-\sum_c \mu_c{\rm N}_c)}}.
\end{equation}
The normalization is explicitly accounted for by the denominator (partition function). 
The second condition means that the energy of a system in heat contact with a thermostat 
fluctuates around an averaged value $\langle {\rm H}\rangle = U=u \Omega$ with the given density of internal energy $u$. 
This condition is taken into account by the Lagrange multiplier $\beta$ that must be related to the temperature,
a more detailed discussion leads to $\beta = 1/(k_{\text{B}}T)$. 
Similar, the contact with the particle reservoir fixes the particle density $n_c$, introduced by the Lagrange multiplier $\mu_c$ that represent the chemical potential of the species $c$.

Within the variational approach, the Lagrange parameters $\beta, \mu_c$ have to be eliminated. 
This leads to the equations of state ($\langle \dots \rangle_{\rm eq}={\rm Tr} \{ \rho_{\rm eq} \dots \}$) which relate, e.g., the chemical potentials $\mu_c$ to the particle densities $n_c$,
\begin{equation}
\langle {\rm H} \rangle_{\rm eq} = U(\Omega, \beta, \mu_c),\qquad 
\langle {\rm N}_c \rangle_{\rm eq} = \Omega n_c(T, \mu_c)\,.
 \end{equation}
The entropy $S_{\rm eq}(\Omega,\beta,\mu)$ follows from Eq. (\ref{Seq}). 
The dependence of extensive quantities on the volume $\Omega$ is trivial for homogeneous systems.
After a thermodynamic potential is calculated, all thermodynamic variables are derived in a consistent manner.
The method to construct statistical ensembles from the maximum of entropy at given conditions, 
which  take into account the different contacts with the surrounding bath, 
is well accepted in equilibrium statistical mechanics and is applied successfully to different phenomena, including phase transitions. 

Can we extend the definition of equilibrium entropy (\ref{Seq}) also for $ \rho(t) $ which describes the evolution in nonequilibrium?
Time evolution is given by an unitary transformation that leaves the trace invariant. 
Thus the expression $ {\rm Tr} \{\rho(t) \ln \rho(t)\}$ is constant for a solution $\rho(t)$ of the von Neumann equation,
\begin{equation}
\label{seqt}
 \frac{\dif}{\dif t} \left[{\rm Tr} \{\rho(t) \ln \rho (t) \} \right]= 0.
\end{equation}
The entropy for a system in nonequilibrium, however, may increase with time according to the second law of thermodynamics.
The equations of motion, including the Schr\"odinger equation and the Liouville-von Neumann equation, 
describe reversible processes and are not appropriate to describe irreversible processes. 
Therefore, the entropy concept (\ref{Seq}) elaborated in equilibrium statistical physics together with the Liouville-von Neumann equation
cannot be used as fundamental approach to nonequilibrium statistical physics.\\

{\it The relevant statistical operator}.
A solution of the problem to combine equilibrium thermodynamics and non-equilibrium processes was proposed by Zubarev \cite{Zubarev}. To characterize the nonequilibrium state of a system, we  introduce the set of relevant observables $\{ {\rm B}_n\}$ extending the set of conserved quantities $\{ {\rm C}_n\}$. At time $t$, the observed values $\langle {\rm B}_n \rangle^t$ have to be reproduced by the statistical operator $\rho(t)$, i.e. 
\begin{equation}
\label{Bnt}
 {\rm Tr} \{\rho(t)\, {\rm B}_n\} = \langle {\rm B}_n \rangle^t
\end{equation}
However, these conditions are not sufficient to fix $\rho(t)$, and we need an additional principle to find the correct one in between many possible distributions which all fulfill the conditions (\ref{Bnt}). In a first step, we can ask for the most probable distribution at time $t$ where the information entropy has a maximum value,
\begin{equation}
-\delta  \left[{\rm Tr} \{\rho_{\rm rel}(t) \ln \rho_{\rm rel} (t) \} \right]= 0
\end{equation}
with the self-consistency conditions
\begin{equation}
\label{selfconsistent}
 {\rm Tr} \{\rho_{\rm rel}(t) {\rm B}_n\} = \langle {\rm B}_n \rangle^t
\end{equation}
and $ {\rm Tr} \{\rho_{\rm rel}(t)\} = 1$. Once more, we use Lagrange multipliers $\lambda_n(t)$ to account for the 
self-consistency conditions (\ref{selfconsistent}).
Since the averages are in general time dependent, the corresponding Lagrange multipliers 
are now time dependent functions as well. 
We find the
generalized Gibbs distribution
	\begin{equation}
	\label{ch4030}
	\displaystyle
	\rho_{\text{rel}}(t)=\exx{-\Phi(t)-\sum\limits_n\lambda_n(t)\op{B}_n},
	\qquad
	\Phi(t)=\ln\,\,{\rm Tr}\,\left\{\exx{-\sum\limits_n\lambda_n(t)\op{B}_n}\right\},
	\end{equation}
where the Lagrange multipliers $\lambda_n(t)$ (thermodynamic parameters) are
determined by the self-consistency conditions
\begin{equation}
\label{ch4040}
{\rm Tr}\,\,\{\rho_{\text{rel}}(t)\op{B}_n\}\equiv\langle \op{B}_n\rangle_{\text{rel}}^t=\langle \op{B}_n\rangle^t.
\end{equation}
$\Phi (t)$ is the Massieux-Planck function, needed for normalization purposes 
and playing the role of a thermodynamic potential.
Generalizing the equilibrium case, Eq. (\ref{Seq}), we can consider the {\it relevant entropy in nonequilibrium}
\begin{eqnarray}
\label{Srel}
S_{\text{rel}}(t)&=&-k_{\text{B}}\,\,\op{Tr}\,\,\{\rho_{\text{rel}}(t)\,\,\ln\rho_{\text{rel}}(t)\}\,.
\end{eqnarray}
Relations similar to the relations known from equilibrium  thermodynamics can be derived. In particular, 
the production of entropy results as
	\begin{equation}
	\label{ch4110}
	\displaystyle
	\frac{\partial S_{\text{rel}}(t)}{\partial t}=\sum\limits_n\lambda_n(t)\langle \op{\dot{B}}_n\rangle^t.
	\end{equation}
as known from the thermodynamics of irreversible processes. In contrast to Eq. (\ref{seqt}), this expression 
can have a positive value so that $S_{\text{rel}}(t)$ can increase with time.

The relevant statistical operator $\rho_{\text{rel}}(t)$ is not the wanted nonequilibrium statistical operator $\rho(t)$ because it does not obey the Liouville-von Neumann equation. Also, $S_{\text{rel}}(t)$ is not the thermodynamic entropy because it is based on the arbitrary choice of the set $\{ {\rm B}_n\}$ of relevant observables, and not all possible variables are correctly reproduced. As example we consider below the famous Boltzmann entropy which is based on the single particle distribution function, but does not take into account higher order correlation functions
(see also the discussion in Sec. \ref{sec:NSO}). 

There are systems in nonequilibrium which are frozen-in, i.e. some degrees of freedom are changing very slowly. For instance, explosives (like oxyhydrogen or dynamite) where the reaction rate is nearly zero are metastable and can be considered as nearly equilibrium states, only with the composition as additional relevant observable characterizing the nonequilibrium  state. Any slow variable may be considered as relevant observable. For instance, in dense, strongly interacting systems we can assume local thermodynamic equilibrium characterized by the densities of conserved quantities, and the gradients of these densities determine the corresponding currents as described by the thermodynamics of irreversible processes. The relevant statistical operator $\rho_{\text{rel}}(t)$ is only an approximation for the nonequilibrium statistical operator $\rho(t)$ which, for appropriate conditions, may reproduce many signatures of the true nonequilibrium state. (Strictly speaking, the concept of conserved quantities often turns out to be an idealization neglecting slow processes like (nuclear) reactions in our normal environment, and the success of equilibrium thermodynamics in real situations is not caused by the rigorous constancy in time of the conserved quantities.) \\

{\it The Zubarev solution of the initial value problem}.
The solution of the problem how to find the missing signatures of $\rho(t)$ not already described by  $\rho_{\text{rel}}(t)$ was found by Zubarev \cite{Zubarev} generalizing the Bogoliubov principle of weakening of initial correlations \cite{Bogoliubov}. 
He proposed to use the relevant statistical operator $\rho_{\text{rel}}(t_0)$ at some initial time $t_0$ 
as initial condition to construct  $\rho(t)$,
\be
	\label{2.205a}
	\begin{array}{rcl}
	\rho_{t_0}(t)=\op{U}(t,t_0)\rho_{\text{rel}}(t_0)\op{U}^\dagger(t,t_0).
	\end{array}
	\ee
The unitary time evolution operator $\op{U}(t,t_0)$ is the solution of the differential equation
\be
\label{2.14}
\ii \hbar\frac{\partial }{\partial t}\op{U}(t,t_0)=\Hop^t\op{U}(t,t_0)\,,
\ee
with the initial condition $\op{U}(t_0,t_0)=1$. This unitary operator is known from the solution of the Schr\"odinger equation.
If the Hamiltonian is not time dependent, we have 
\begin{equation}
\op{U}(t,t_0) = {\rm e}^{- \frac{\ii}{\hbar}\Hop (t-t_0)}\,.
\end{equation}
If the  Hamiltonian is time dependent, the solution is given by a time-ordered exponent.

Now, it is easily shown that $\rho_{t_0}(t)$ is a solution of the von Neumann equation. 
All missing correlations not contained in $\rho_{\text{rel}}(t_0)$ are formed dynamically during the time evolution of the system. 
However, incorrect initial correlations contained in  $\rho_{\text{rel}}(t_0)$ may survive for a finite time interval $t-t_0$, and the self-consistency conditions (\ref{selfconsistent}) valid at $t_0$ are not automatically valid also at $t$. 

To get rid of these incorrect initial correlations, according to the Bogoliubov principle of weakening of initial correlations one can consider the limit $t_0 \to -\infty$. 
According to Zubarev, it is more efficient to average over the initial time so that
no special time instant $t_0$ is singled out. This is of importance, for instance, if there are long living oscillations determined by the initial state.
According to Abel's theorem, see Refs. \cite{Zubarev,ZMR1,ZMR2}, the limit  $t_0 \to - \infty$ can be replaced by the limit $\epsilon \to +0$ in the expression
\be
	\label{2.205}
	\begin{array}{rcl}
	\rho_{\epsilon}(t)=\epsilon\int\limits_{-\infty}^t\exx{\epsilon(t_1-t)}
	\op{U}(t,t_1)\rho_{\text{rel}}(t_1)\op{U}^\dagger(t,t_1)\dif t_1.
	\end{array}
	\ee
This averaging over different initial time instants means a mixing of phases so that long-living oscillations are damped out. Finally we obtain the nonequilibrium statistical operator as
\be
	\label{2.206}
	\rho_{\text{NSO}}(t)=\lim_{\epsilon\rightarrow 0}\rho_{\epsilon}(t)\,.
	\ee

This way,  $\rho_{\text{rel}}(t_1)$ for  all times $-\infty <t_1<t$ serves as initial condition to solve the  Liouville-von Neumann equation
according to the Bogoliubov principle of weakening of initial correlations. The missing correlations are formed dynamically during the time evolution of the system.
The more information about the nonequilibrium state are used to construct the relevant statistical operator, the less dynamical formation
of the correct correlations in $ \rho(t)$ is needed. The limit $t_0 \to - \infty$ is less active to produce the remaining missing correlating. The past that is of relevance, given by the relaxation time $\tau$, becomes shorter, if the relevant (long-living) correlations are already correctly implemented. 
One one hand, the limit  $\epsilon \to +0$ is to be considered as $\epsilon \ll 1/\tau$.
The limit  $\epsilon \to +0$ has to be performed after the thermodynamic limit, see the following section, topic 3.

\section{Discussion of the Zubarev NSO expression}
\label{sec:NSO}

{\it The extended Liouville-von Neumann equation}.
The nonequilibrium statistical operator $\rho_{\epsilon}(t)$, Eq. (\ref{2.205}), obeys the  extended von Neumann equation	
\be
\label{vNNSO}
		\frac{\partial\rho_{\epsilon}(t)}{\partial t}+\frac{\ii}{\hbar}[\Hop^t,\rho_{\epsilon}(t)]=-\epsilon(\rho_{\epsilon}(t)-\rho_{\text{rel}}(t)).
	\ee
as can be seen after simple derivation with respect to time. In contrast to the von Neumann equation (\ref{1.5}), 
a source term arises on the right hand side that becomes infinitesimal small in the limit $\epsilon \to +0$. This source term breaks the time inversion symmetry so that, for any finite value of $\epsilon$, the solution $\rho_{\epsilon}(t)$ describes in general an irreversible evolution with time.

The source term can be interpreted in the following way:
\begin{enumerate}
\item
The source term implements the 'initial condition' in the equation of motion as expressed by $\rho_{\text{rel}}(t)$. 
Formally, the source term looks like a relaxation process. In addition to the internal dynamics, 
the system evolves towards the relevant distribution.

\item
The construction of the source term is such that the time evolution of the relevant 
variables is not affected by the source term (we use the invariance of the trace 
with respect to cyclic permutations),
\be
		\frac{\partial}{\partial t} \langle {\rm B}_n \rangle^t=
{\rm Tr} \left\{\frac{\partial\rho_{\epsilon}(t)}{\partial t}{\rm B}_n \right\}=
-{\rm Tr} \left\{\frac{\ii}{\hbar}[\Hop^t,\rho_{\epsilon}(t)]{\rm B}_n \right\}
= \left\langle \frac{\ii}{\hbar}[\Hop^t,{\rm B}_n] \right\rangle^t=\langle \dot {\rm B}_n \rangle^t\,.
\ee
The source term cancels because of the self-consistency conditions (\ref{selfconsistent}).
 Thus, the time evolution of the relevant observables satisfies the dynamical equations of motion according to the Hamiltonian $\Hop^t$.

\item
The value of $\epsilon$ has to be small enough,  $\epsilon \ll 1/\tau$, so that all relaxation processes to establish the correct correlations,
i.e. the correct distribution of the irrelevant observables, can be performed. However, $\hbar \epsilon$ has to be large compared to the energy difference of neighbored energy eigenstates of the system so that mixing is possible. For a system of many particles, the density of energy eigenvalues is high so that we can assume a quasi-continuum. This is necessary to allow for dissipation. The van Hove limit means that the limit  $\epsilon \to +0$ has to be performed after the thermodynamic limit.

\item
Differential equations can have degenerated solutions. For instance, we know the retarded and advanced solution of the wave equation which describes the emission of electromagnetic radiation. An infinitesimal small perturbation can destroy this degeneration and select out a special solution, here the retarded one. Similar problems are known for systems where the ground state has a lower symmetry than the Hamiltonian. Examples are the spontaneous magnetization below the Curie point of a Heisenberg model ferromagnet that breaks isotropy, or, at the liquid-solid phase transition, the formation of a lattice in crystals that breaks homogeneity of space. 

\item
Any real system is in contact with the surroundings. The intrinsic dynamics described by the Hamiltonian $\Hop^t$ is modified due to the coupling of the open system to the bath. Within the quantum master equation approach, we can approximate the influence term describing the coupling to the bath by a relaxation term as given by the source term. We come back to this issue below in Sec. \ref{Outlook}. 
However, at present we consider the source term as a purely mathematical tool to select the retarded solution of the 
Liouville-von Neumann equation, and physical results are obtained only after performing the limit $\epsilon \to 0$.
\end{enumerate}

{\it Selection of the set of relevant observables}.
The Zubarev method  to solve the initial value problem for the Liouville-von Neumann equation is based on the selection 
of the set $\{{\rm B}_n\}$ of relevant observables which characterize the nonequilibrium state. 
The corresponding relevant statistical operator $\rho_{\rm rel}(t)$ is some approximation to $\rho(t)$.
According to the Bogoliubov principle of weakening of initial correlations, the missing correlations to get $\rho(t)$ are produced dynamically.
This process, the dynamical formation of the missing correlations, needs some relaxation time $\tau$.
If we would take instead of  $\rho_{\rm rel}(t)$ the exact (but unknown) solution $\rho(t)$, the relaxation time $\tau$ is zero.
The Liouville-von Neumann equation, which is a first order differential equation with respect to time, describes a Markov process.

There is no rigorous prescription how to select the set of relevant observables $\{{\rm B}_n\}$. 
The more relevant observables are selected so that their averages with  $\rho_{\rm rel}(t)$ reproduce already the correctly
known averages $\langle {\rm B}_n \rangle^t$, see Eq.~(\ref{selfconsistent}), 
the less the effort to produce the missing correlations dynamically, and the less relaxation time $\tau$ is needed.
Taking into account that usually perturbation theory is used to treat the dynamical time evolution (\ref{2.14}), 
 a lower order of perturbation theory is then sufficient.

In conclusion, the selection of the set of relevant observables is arbitrary, as a minimum the constants of motion ${\rm C}_n$
have to be included because their relaxation time is infinite, their averages cannot be produced dynamically. 
The resulting $\rho_{\text{NSO}}(t)$ (\ref{2.206}) should not depend on the (arbitrary) choice of relevant observables $\{{\rm B}_n\}$
if the limit $\epsilon \to 0$ is correctly performed.
However, usually perturbation theory is applied, so that the result will depend on the selection of the set of relevant observables.
The inclusion of long-living correlations into $\{{\rm B}_n\}$ allows to use lower order perturbation expansions to obtain acceptable results. We come back to this in Sec. \ref{Operatorset}.

We consider the electrical conductivity of charged particle systems below in Sec. \ref{conductivity}. 
As relevant observables we will consider the following sets, see Refs. \cite{Roep88,r90,Redmer97,Reinholz05,rr12}: \\
1. only the constants of motion, e.g. particle number and energy (cf. Eq. (\ref{gr.can})),\\
2. the particle currents as known from the thermodynamics of irreversible processes,\\
3. the single-particle distribution function as known from kinetic theory, and\\
4. two-particle correlation functions needed to describe interacting systems.\\
Different results are obtained if different sets of relevant observables are selected as well 
as low order perturbation expansion is performed.
It is expected that the same result will appear after summing up all orders of the perturbation expansion, independent of the 
choice of the set of relevant observables $\{{\rm B}_n\}$.\\

{\it Entropy of the nonequilibrium state}.
An intricate problem is the definition of entropy for the nonequilibrium state. 
In nonequilibrium, entropy is produced, as investigated in the phenomenological approach to the thermodynamics of irreversible processes,
considering currents induced by the generalized forces.

Such a behavior occurs for the relevant entropy defined by the 
relevant distribution (\ref{Srel}), 
\begin{equation}
S_{\text{rel}}(t)=-k_{\text{B}} {\rm Tr} \left\{ \rho_{\text{rel}}(t) 
\ln \rho_{\text{rel}}(t) \right\}.
\end{equation}
A famous example that shows the increase of the relevant entropy with time is the 
Boltzmann H (capital Eta) theorem where the relevant observables to define 
the nonequilibrium state is the single particle distribution function. 
Using the Boltzmann equation with the Sto{\ss}zahlansatz, 
it can be shown that the relevant entropy (Boltzmann entropy) based on the 
single-particle distribution function is  increasing with time or remains constant 
for the  equilibrium solution. 
However, the equilibrium solution is the ideal gas what proves that this entropy 
concept is not correct because the contribution of interaction to the well-defined 
entropy in equilibrium is not reproduced.

Note that the increase of entropy cannot be solved this way. 
It is related to so-called coarse graining. 
The information about the state is reduced because 
the degrees of freedom to describe the system are reduced. 
This may be an averaging in phase space over small cells 
or any borderline between a macroscopic observable and a microscopic dynamical state. 
Also the average over different phases, the destruction of quantum interference (dephasing),
 and other projection techniques will destroy information. 
The loss of information then gives the increase of entropy. 
This procedure is artificial, depending on our way to describe the details of a process, 
anthropomorphic, related to our technical possibilities to prepare and measure the state 
of a system and control the dynamics. 
There is no first principle approach that gives the decision how the 
relevant degrees of freedom have to be selected out. 
Of course, in certain situations the choice of relevant observables becomes quite natural, 
we will see this in the following on the case of Quantum master equations (QME), 
see Refs. \cite{Gocke,Cheng}, Kinetic theory (KT), see Ref. \cite{rr12}, and Linear response theory (LRT), see Ref. \cite{ChrisRoep85}. 
From a fundamental point of view, this situation is unsatisfactory.

The method of nonequilibrium statistical operator $\rho_{\text{NSO}}(t) $ allows to extend the set of relevant observables arbitrarily so that
the choice of the set of relevant observables seems to be irrelevant. All missing correlations are produced dynamically. We can start with any set of relevant operators, but have to wait for a sufficient long time to get the correct statistical operator, or to go to very small $\epsilon$. A possible definition of the entropy would be 
\be
S_{\text{NSO}}(t) = -k_B {\rm Tr} \left\{ \rho_{\text{NSO}}(t) \ln \rho_{\text{NSO}}(t) \right\}\,.
\ee
The destruction of the reversibility of the von Neumann equation (\ref{vNNSO}) is connected with the source term on the right hand side that produces the mixing by averaging over the past in Eq. (\ref{2.205}). This source term is responsible for the entropy production. There is at present no proof that the entropy $S_{\text{NSO}}(t)$ will increase also in the limit  $\epsilon \to +0$. A fundamental process causing the production of entropy is missing in the approaches given here.

\section{Generalized linear response theory}

{\it Response to an external field}.
We consider a system under the influence of  
external (time dependent) fields acting on the particles, see \cite{ZMR2,rerrw00,rr12},
\begin{equation}
\Hop^t=\Hop_{\rm S}+\Hop_{\rm F}^t,
\end{equation}
where $\Hop_{\rm S}$ denotes the system Hamiltonian, containing all kinetic energies of 
the particles as well as the full interaction part. The second part $\Hop_{\rm F}^t$ 
describes the coupling of the system to the external fields $h_j$: 
\begin{equation}
\label{ch6020}
\Hop_{\rm F}^t=-\sum_jh_j\exx{-\ii\omega t}\op{A} _j.
\end{equation}
We consider the limit of weak external fields. Compared with the equilibrium distribution (\ref{gr.can}) we expect that 
the changes of the state of the system are also weak. We characterize the nonequilibrium state by the set  $\{{\rm B}_n\}$ of relevant observables
and assume that the averages
\begin{equation}
\label{ch6020a}
\langle \op{B}_n \rangle^t={\rm Tr}\{\rho(t) \op{B}_n \} \propto h_j
\end{equation}
are proportional to the external fields (linear response). In the following we assume that the equilibrium expectation values of the nonequilibrium fluctuations disappear, $\langle\op{B}_n\rangle_{\rm eq}=0$ (else we have to subtract the equilibrium values).
 
Treating the conserved observables explicitly, we write the relevant statistical operator $\rho_{\rm rel}$ 
in the form ($\mathcal H=\Hop_{\rm S}-\sum_c\mu_c\op{N}_c$)
\bea
\label{ch60310}
\rho_{\rm rel}(t)
&=&\exx{-\Phi(t)-\beta\left(\mathcal H-\sum\limits_nF_n(t)\,\op{B}_n\right)},\qquad \Phi(t)=\ln {\rm Tr}\,\left\{\exx{-\beta\left(\mathcal H-\sum\limits_nF_n(t)\,\op{B}_n\right)}\right\}\,,
\eea
where the Lagrange multipliers are divided into the equilibrium parameters $\beta, \mu$
and  the generalized response parameters $F_n(t)$, coupled to the corresponding observables.
All Lagrange parameters are determined by the given mean values of these observables.
In particular, we have  the self consistency conditions (\ref{selfconsistent})
\begin{equation}
\label{036}
\langle \op{B} _n\rangle_{\rm rel}^t={\rm Tr}\,\{\rho_{\rm rel}(t)\op{B} _n\}={\rm Tr}\,\{\rho(t)\op{B} _n\}=\langle \op{B} _n\rangle^t
\end{equation}
or
\begin{equation}
\label{036a}
{\rm Tr}\,\{\rho_{\rm irrel}(t)\op{B} _n\}=0, \qquad  \rho_{\rm irrel}(t) = \rho(t)-\rho_{\rm rel}(t)\,.
\end{equation}

The corresponding self consistency condition for  $\op{N}$ and $\Hop_{\rm S}$ lead to the well-known equations of state for the temperature $1/\beta$ and the chemical potential $\mu$.
$\Phi(t)$ is the Massieu-Planck functional that normalizes $ \rho_{\rm rel}(t) $.

The basic assumption of LRT is that the average values $\langle \op{B}_n \rangle^t$ of the additional observables, which characterize the response of the system, are proportional to the external fields. 
Because these external fields are arbitrarily weak, we expand all quantities with respect to the
fields up to first order. If the fluctuations $\langle \op{B}_n \rangle^t$ are
proportional to these fields, we have also  $F_n\propto h_j$. Below we derive linear 
equations that relate the response of the system to the causing external fields.

In the linear regime we await the response parameters $F_n(t)$ to
exhibit the same time dependence as the external fields:
\be
\label{ch60320}
F_n(t)=F_n\exx{-\ii\omega t}.
\ee
Here we have harmonic fields $h_j \exx{\ii\omega t}$, but the formulation rests 
general as we can always express arbitrary time dependences by means of a
Fourier transformation. Within the linear regime, the superposition of different components of the field gives the 
superposition of the corresponding responses. The treatment of spatial dependent external forces is also
possible. 
As a specific advantage of the Zubarev method,  thermodynamic forces such as gradients of temperature or 
chemical potentials can be treated \cite{ZMR2,r90,Redmer97,Reinholz05}.\\

{\it Elimination of the Lagrange multipliers}.
The main problem is to eliminate the Lagrange multipliers, the generalized response parameters $F_n(t)$. This is possible explicitly in the case of kinetic theory (KT), and this is also possible explicitly in the case of linear response
theory (LRT).
With the operator relation
	\be
	\label{001}
	\exx{\op{A} +\op{B} }=\exx{\op{A} }+\int\limits_0^1 \mathrm{d}\lambda\, \exx{\lambda (\op{A} +\op{B} )}\op{B} \,\exx{(1-\lambda)\op{A} }.
	\ee\\
we get for the relevant statistical operator (\ref{ch60310}) up to first order of the nonequilibrium fluctuations $\{\op{B}_n\}$
	\be
	\label{rellin}
	\rho_{\rm rel}(t)= \rho_{\rm eq}+\beta\int\limits_0^1\mbox{d}\lambda \sum_n F_n(t)\,\op{B}_n(\ii\hbar\beta\lambda)\, \rho_{\rm eq}.
	\ee
Here we made use of the modified Heisenberg picture $\op{O}(\tau)=\exp(\ii {\mathcal H} \tau/\hbar) \op{O}\exp(-\ii {\mathcal H} \tau/\hbar)$
 with $\tau\to\ii\hbar\beta\lambda$
replacing in the exponents $\Hop_{\rm S}$ by	$\mathcal H=\op{H}_{\rm S}
-\sum_c\mu_c\op{N}_c$. 
We want to calculate expectation values of macroscopic
relevant variables that commute with the particle number operator $\op{N}_c$ so that
we can use both $\mathcal H$ and $\op{H}_{\rm S}$ synonymously.
(Mention that also the Massieu-Planck functional $\Phi(t)$ has to be expanded so that the fluctuations around the equilibrium averages $\{{\rm B}_n-\langle {\rm B}_n \rangle_{\rm eq}  \}$ appear.)

After linearization with respect to the external fields $h_j$ and the response parameters $F_n$, see appendix \ref{App:1}, finally we have 
\bea\nn
\rho_\epsilon(t)&=&\rho_{\rm rel}(t)-\beta\,\exx{-\ii\omega t}
\int\limits_{-\infty}^0\dif t_1\,\exx{-\ii z t_1}\int\limits_0^1\dif\lambda\left[-\sum\limits_j h_j\,\dot{\op{A}}_j(\ii\lambda\beta\hbar+t_1)\,\rho_{\rm eq}\right.\\\label{rhoErgebnis}
&&\left.+\sum\limits_n\left(F_n\,\dot{\op{B}}_n(\ii\lambda\beta\hbar+t_1)\,\rho_{\rm eq}-\ii\omega F_n\,\op{B}_n(\ii\lambda\beta\hbar+t_1)\,\rho_{\rm eq}\right)\right]
\eea
($z=\omega+\ii\epsilon$). Here we used that $h_j(t)$ and $F_n(t)$, Eq. (\ref{ch60320}), are proportional to $\mbox{e}^{-\mbox{i}\omega t}$.

We multiply this equation by $\op{B} _m$, take the trace and use the 
self consistency relation (\ref{036}). We obtain a set of 
linear equations for the thermodynamically conjugated parameters $F_n$ (response
parameters): 
	\begin{equation}
	\label{ch6200}
	\sum_n\left\{\langle\op{B}_m; \dot{\op{B} }_n\rangle_z
	-\ii\omega\langle\op{B}_m;  \op{B} _n \rangle_z\right\}F_n=
	\sum_j\langle\op{B}_m; \dot{\op{A} }_j\rangle_zh_j,
	\end{equation}
with the Kubo scalar product (the particle number commutes with the observables)
\be
(\op{A}\,|\,\op{B})=\int\limits_0^1\dif \lambda \op{Tr}\,\left\{\op{A}\, \exx{-\lambda \beta \cal{H}}\,
\op{B}\, \exx{\lambda \beta \cal{H}}\,\rho_{\rm eq}\right\}=\int\limits_0^1\dif \lambda \,\op{Tr}\,\left\{\op{A}\,
\op{B}(\ii\lambda\beta\hbar)\,\rho_{\rm eq}\right\},
\ee
and its Laplace transform, the thermodynamic correlation function
\bea
\langle \op{A};\op{B} \rangle_z&=&\int\limits_{-\infty}^0 \mathrm{d}t\, \exx{-\ii zt}(\op{A}\,|\,\op{B}(t) )=
\int\limits_0^{\infty}\dif t\, \exx{\ii zt}(\op{A}(t)\,|\,\op{B}).
\eea

The linear system of equations (\ref{ch6200}) has the form
\begin{equation}
	\label{GLRE1}
	\sum_n P_{mn}F_n=\sum_j D_{mj}h_j
\end{equation}
to determine the response parameters $F_n$, the number of equations coincides with the number of variables to be determined. The coefficients of this linear system of equations are given by equilibrium correlation functions. We emphasize that in the classical limit the relations become more simple because the variables commute, and we have not additional integrals expanding the exponential.

We can solve this linear system of equations (\ref{ch6200}) using Cramers rule. The response parameters  $F_n$ are
found to be proportional to the external fields $h_j$ with coefficients that are ratios of two determinants.
The matrix elements are given by equilibrium correlation functions. This way, the self-consistency conditions 
are solved, and the Lagrange multipliers can be eliminated. The non-equilibrium problem is formally
solved. The second problem, the evaluation of equilibrium correlation functions, can be solved by different
methods such as numerical simulations, quantum statistical perturbation theories such as thermodynamic Green functions and Feynman diagrams, path integral methods, etc. 
Using  partial integration, we show the relation 
\bea
-\ii z\langle \op{A};\op{B}\rangle_z
\label{ch6240}
&=& (\op{A}\,|\,\op{B})+\langle \dot{\op{A}};\op{B}\rangle_z=
(\op{A}\,|\,\op{B})-\langle \op{A};\dot{\op{B}}\rangle_z.
\eea
Then, the generalized linear response equations (\ref{ch6200}) can be rewritten  
in the short form (\ref{GLRE1}) with the matrix elements
\bea
\label{GLRE2}
P_{mn}&=&(\op{B}_m|\dot{\op{B}} _n)+\langle \dot{\op{B} }_m;\dot{\op{B} }_n\rangle_{\omega+\ii\epsilon}-\ii\omega(\op{B} _m|\op{B} _n)-\ii\omega\langle \dot{\op{B} }_m;\op{B}_n\rangle_{\omega+\ii\epsilon}\\
D_{mj}&=&(\op{B}_m|\dot{\op{A}}_j)+\langle\dot{\op{B} }_m;\dot{\op{A} }_j\rangle_{\omega+\ii\epsilon}.
\eea
that can be interpreted as generalized transition rates (collision integral, left hand side) and the influence 
of external forces (drift term, right hand side).

Having the response parameters $F_n$ to our disposal, we can evaluate averages of the
relevant observables, see Eq. (\ref{036}),
\begin{equation}
\label{ch6250}
\langle \op{B} _n\rangle^t=\langle \op{B} _n\rangle\ind{rel}^t=-\beta \sum_m F_m\exx{\ii\omega t}N_{mn},
\qquad
N_{mn}=(\op{B} _m|\op{B} _n).
\end{equation}
Eliminating $F_n$, these average fluctuations $ \langle \op{B} _n\rangle^t$ are 
proportional to the fields $h_j$.

\section{Conductivity of a plasma}\label{conductivity}

{\it Force-force correlation function and static (dc) conductivity}.
As an example for the generalized linear response theory, we calculate the 
conductivity of a plasma of charged particles (electrons and ions) 
that is exposed to a static 
homogeneous electric field in $x$-direction: $\omega = 0$, \,\,$\bs{E}=(E,0,0) = E \bs e_x$, 
\begin{equation}
\label{4:6}
\Hop_{\rm F}=-eE\op{X},
\qquad
\op{X}=\sum_i^{N_e}\op{x}_i.
\end{equation}
Instead of $h_j$ we have only one constant external field $E$. For the treatment of arbitrary $\omega$ to obtain 
the dynamical (optical) conductivity see Refs. \cite{Roep98,rerrw00,Reinholz05,rr12}. The conjugated variable $\op{A} $ from
\Eq{ch6020} that couples the system to the external field is $\op{A} =e\op{X}$. 
The time derivative follows as $\dot{\op{A}}=(e/m) \op{P}$, with 
$\op{P}=\sum_i^{N_e}\op{p}_{x,i}$ denoting the total momentum in $x$ direction.

For simplicity, the ions are considered here as fixed in space because of the large mass ratio (adiabatic approximation).  
Then, the transport of charge is owing to the motion of the electrons.
In general, the ions can also be treated as moving charged particles that contribute to the current.

A stationary state will be established in the plasma where the electrons are accelerated by the external field,
but loose energy (and momentum) due to collisions with the ions. This nonequilibrium state is characterized 
by an electrical current that is absent in thermal equilibrium. 
We can take  the electric current density $\op{j}_{\rm el} = (e/m \Omega) \op{P} = (e/\Omega)
\dot{\op{X}}$ as a relevant observable that characterizes the nonequilibrium state. 
Instead, we take the total momentum  $\op{B}=\op{P} =m\dot{ \op{X}}$.
The  generalized linear response equations (\ref{GLRE1}), (\ref{GLRE2}) read
\begin{equation}
\label{ch6300}
F\left[(\dot{\op{P}}|{\op{P}})+\langle \dot{\op{P}};\dot{\op{P}}\rangle_{\ii\epsilon}\right]=\frac{e}{m}E\{(\op{P}|\op{P})+
\langle \op{P};\dot{\op{P}}\rangle_{\ii\epsilon}\},
\end{equation}
The term $(\dot{\op{P}}|{\op{P}})$ vanishes as can be shown with
	Kubo's identity, see \Eq{003},
	\be
	\label{ch6320}
	(\dot{\op{P}}|\op{P})=\langle[\op{P},\op{P}]\rangle_{\rm eq}=0.
	\ee
With the Kubo identity, we also evaluate the Kubo scalar product
\begin{equation}
\label{ch6380}
(\op{P}|\op{P})=m \int\limits_0^1\dif\lambda\langle\dot{\op{X}}
(-\ii\hbar\beta\lambda)\op{P}\rangle_{\rm eq}=-\frac{\ii \,m}{\hbar\beta}
\op{Tr}\left\{\rho_{\rm eq}[\op{X},\op{P}]\right\}
=\frac{mN}{\beta}.
\end{equation}

The solution for response parameter $F$ is
\begin{equation}
F=\frac{e}{m} E \frac{\frac{mN}{\beta}+ \langle\op{P};\dot{\op{P}}\rangle_{\ii\epsilon}}{\langle\dot{\op{P}};\dot{\op{P}}\rangle_{\ii\epsilon}}.
\end{equation}

With \Eq{ch6250} we have
\bea
\label{ch6330}
\langle {\rm j}_{\rm el}\rangle&=&\frac{e}{m\Omega}\langle\op{P}\rangle_{\rm rel}=\frac{e\beta}{m\Omega}\,F\,(\op{P}|\op{P})=\sigma_{\rm dc} E\,.
\eea
The resistance $R$ in the static limit follows as
	\begin{equation}
	\label{ch6350}
	R=\frac{1}{\sigma_{\rm dc}}=\frac{\Omega \beta}{e^2N^2}\frac{\langle\dot{\op{P}};\dot{\op{P}}\rangle_{\ii\epsilon}}{1+\frac{\beta}{mN}
	\langle\op{P};\dot{\op{P}}\rangle_{\ii\epsilon}}.
	\end{equation}

{\it Ziman formula for the Lorentz plasma}.
To evaluate  the resistance $R$ we have to
calculate the correlation functions $\langle\dot{\op{P}};\dot{\op{P}}\rangle_{\ii\epsilon}$ 
and $\langle\op{P};\dot{\op{P}}\rangle_{\ii\epsilon}$. For this we have to specify the system Hamiltonian 
$\Hop_{\rm S}$, which reads for the Lorentz plasma model
	\begin{equation}
	\label{ch6370}
	\Hop_{\rm S}=\Hop_{0}+\Hop_{\rm int}=\sum_{\bs p}E_{\bs p}
\aopd_{\bs p}\aop_{\bs p}+\sum_{\bs p,\bs q}V_q\aopd_{\bs p+\bs q}\aop_{\bs p}\,,
\qquad E_{\bs p}=\frac{\hbar^2p^2}{2m}.
	\end{equation}
We consider the ions at fixed positions ${\bs R}_i$ so that $V({\bs r})=\sum_i V_{\rm ei}({\bs r}-{\bs R}_i)$.
The Fourier transform $V_q$ depends for isotropic systems only on the modulus $q=|{\bs q}|$ 
and will be specified below.
 A realistic plasma Hamiltonian should consider also moving ions and the electron-electron interaction 
so that we have a two component plasma Hamiltonian with pure Coulomb interaction between all constituents. 
This has been worked out \cite{Roep88} but is not subject of our present work so that we restrict ourselves mainly
to the simple Lorentz model.

The force $\dot{\op{P}}$ on the electrons follows from
 the $x$ component of the total momentum ($p$ is the wave number vector)
	\be
	\label{ch6371}
	{\rm P}=\sum_{\bs p}\hbar p_x\,\aopd_{\bs p}\aop_{\bs p}.
	\ee
as
	\bea
	[\Hop_{\rm S},{\rm P}]&=&
-\sum_{\bs p,\bs q }V_q\,\hbar q_x\,\aopd_{\bs p+\bs q}
	\aop_{\bs p}
	\eea
We calculate the force-force correlation function
	(only $x$ component)
	\be
	\label{ch6390}
	\langle\dot{\op{P}};\dot{\op{P}}\rangle_{\ii\epsilon}=
	\int\limits_{-\infty}^0\dif t\,\exx{\epsilon t}\int\limits_0^1\dif\lambda
	\left\langle\frac{\ii}{\hbar}[\Hop_{\rm S},\op{P}(t-\ii\lambda\beta\hbar)]\frac{\ii}{\hbar}[\Hop_{\rm S},
	\op{P}]\right\rangle_{\rm eq}
	\ee
in Born approximation with respect to $V_q$. In lowest order, the force--force correlation function is of second order so that
in the time evolution $\exp[(\ii/\hbar) \Hop_{\rm S} (t-\ii\lambda\beta\hbar)]$ the contribution $\Hop_{\rm int}$ of interaction 
to $\Hop_{\rm S}$, Eq. (\ref{ch6370}), can be dropped as well as in the statistical operator.
	The averages are performed with the non-interacting $\rho_0$. 
The product of the two commutators is evaluated using Wick's theorem.  
	One obtains
	\bea\nn
\langle\dot{\op{P}};\dot{\op{P}}\rangle_{\ii\epsilon}&=&-\sum_{\bs p,\bs p',\bs q,\bs q'}\int\limits_{-\infty}^0\dif t\,\exx{\epsilon t}\int\limits_0^1\dif\lambda\,\exx{\frac{\ii}{\hbar}(E_{\bs p}-E_{\bs p+\bs q})(t-\ii\hbar\beta\lambda)}
	V_qV_{q'} q_x q'_x \langle \aopd_{\bs p+\bs q}
	\aop_{\bs p}\aopd_{\bs p'+\bs q'} \aop_{\bs p'} \rangle_{\rm eq}
	\\
	\label{ch6410}
	&=&\sum_{\bs p,\bs q}|V_{ q}|^2\delta(E_{ p}-E_{\bs p+\bs q})f_{ p}(1-f_{ p})\pi\hbar  q_x^2.
	\eea
Because the $x$ direction can be arbitrarily chosen in an isotropic system, we replace $ q_x^2 = ( q_x^2+ q_y^2+ q_z^2)/3= q^2/3$ if the remaining contributions to the integrand are not depending on the 
direction in space.

Evaluating Eq. (\ref{ch6350}) in Born approximation, the correlation function  $\langle\op{P};\dot{\op{P}}\rangle_{\ii \epsilon}(\beta/mN)$ 
can be neglected in relation to 1 because it contains the interaction strength. 
For the resistance, this term contributions 
only in higher orders of the interaction.

The force-force correlation function (\ref{ch6410}) is further evaluated using 
the relations
\be
	-\frac{1}{\beta}\ddif{f(E_{ p})}{E_{ p}}=\frac{\exx{\beta (E_{ p}-\mu)}}{(\exx{\beta (E_{ p}-\mu)}+1)^2}=f_{ p}(1-f_{ p})
	\ee
and
\be
	\label{ch6430}
	\delta (E_{ p} -E_{\bs p+\bs q})=\frac{m}{\hbar^2 qp}\delta (\cos\theta -\frac{q}{2p}),
	\ee
	thus, the $q$ integration has to be performed in the limits 
	$0\leq q\leq 2p$.
Finally the resistance can be calculated by inserting the previous 
expressions \Eq{ch6380} and \Eq{ch6410} into \Eq{ch6350} so that the Ziman-Faber formula is obtained,
\begin{equation}
\label{ch6420}
R=\frac{m^2\Omega^3}{12\pi^3\hbar^3e^2N^2}\int\limits_0^{\infty}\dif E(p)\left(-\frac{\dif f(E)}{\dif E}\right)\int\limits_0^{2p}\dif q\, q^3|V_q|^2.
\end{equation}

The expression for the resistance depends on the special form
of the potential $V_q$. For a pure Coulomb potential 
$e^2/(\Omega \epsilon_0 q^2) $
the integral diverges logarithmically as typical for Coulomb integrals.
The divergency at very small values of $q$ is removed if screening due to
the plasma is taken into account. Within a many-particle approach, in static approximation
the Coulomb potential is replaced by the Debye potential
\be
\label{Debpot}
V_q=\frac{e^2}{\Omega \epsilon_0(q^2+\kappa^2_{\rm D})}
\ee
where $\kappa$ is just the inverse Debye 
screening length, $\kappa^2_{\rm D}=r_{\rm D}^{-2}=\frac{e^2n}{\epsilon_0 k_{\rm B}T}$, and the 
ionic structure factor $\sum_{ij}\e^{\ii{\bf q}\cdot({\bf R}_i-{\bf R}_j)} $ is taken as $N_{\rm ion}$ for uncorrelated ion positions.

We obtain the Coulomb logarithm
\be
\Lambda ( p)=\int\limits_0^{2p}\dif q\, q^3|V_q|^2 =\ln \sqrt{1+b}-\frac{1}{2} \frac{b}{1+b}, \qquad
b=\frac{4 p^2 \epsilon_0}{\beta e^2n_e}
\ee
Performing the low-density limit at fixed temperature, the Fermi distribution function can be replaced by the Boltzmann distribution function. We have
\be
\label{Ziman0}
\sigma_{\rm dc}=\frac{3}{4 \sqrt{2 \pi}} \frac{(k_{\rm B})^{3/2} (4 \pi \epsilon_0)^2}{m^{1/2} e^2 }\frac{1}{\Lambda( p_{\rm therm})}
\ee
where the Coulomb logarithm is approximated by the value of the average $p$, with $\hbar^2 p_{\rm therm}^2/2m=3 k_{\rm B}T/2$.
In the low-density limit, the asymptotic behavior of the Coulomb logarithm $\Lambda$ is given by $-(1/2) \ln n$. However, this result for $\sigma_{\rm dc} $ is not correct and can only be considered as an approximation, as discussed below in following section considering the virial expansion of the resistivity.

\section{Extended set of relevant observables}
\label{Operatorset}

{\it Different sets of relevant observables}. 
After fully linearizing the statistical operator (\ref{rhoErgebnis}) with (\ref{rellin}), we have for the electrical current density 
\bea
\label{PREA15}
\langle {\rm j}_{\rm el} \rangle = \frac{e}{m \Omega}\langle {\rm P} \rangle=\frac{e\beta}{m \Omega}
 \left\{\sum_n\left[({\rm P}|{\rm B}_n)-\langle {\rm P};\dot {\rm B}_n \rangle_{i \epsilon}\right]F_n+\langle {\rm P};{\rm P} \rangle_{i \epsilon}\frac{e}{m} E\right\}= \sigma_{\rm dc} E.
\eea
After deriving the Ziman formula from the force-force correlation function in the previous section, we investigate the question to select an appropriate set of relevant observables $\{ {\rm B}_n\}$.\\

\noindent{\it Kubo formula.}
We consider different choices for the set  $\{{\rm B}_n\}$ of relevant observables. The most simplest case is the empty set. There are no response parameters to be eliminated. According Eq. (\ref{PREA15}), the Kubo formula
\begin{eqnarray}
\label{kubo}
 \sigma^{\rm Kubo}_{\rm dc}&=&
\frac{e^2 \beta}{m^2 \Omega} \langle \op{P} ;\op{P} \rangle_{ \ii \epsilon}^{\rm irred}
\end{eqnarray}
follows \cite{Kubo66}. The index 'irred' denotes the irreducible part of the correlation function, 
because the conductivity is not describing the relation between  the current and the external field, but the internal field. We will not discuss this in the present work. A similar expression can also be given for the dynamical, wave-number vector dependent conductivity $\sigma({\bs q},\omega)$ which is related to other quantities such as the response function, the dielectric function, or the polarization function, see Refs. \cite{Roep98,Reinholz05,rr12,r2}. Eq. (\ref{kubo}) is a fluctuation-dissipation theorem, equilibrium fluctuations of the current density are related to a dissipative property, the electrical conductivity.

The idea to relate the conductivity with the current-current autocorrelation function in thermal equilibrium looks very appealing because the statistical operator is known. The numerical evaluation by simulations can be performed for any densities and degeneracy.
At finite ${\bs q},\omega$, analytical evaluations are possible for noninteracting quantum gases that gives the Random phase approximation (RPA). In the limit ${\bs q} \to 0,\,\,\omega \to 0$, the dc conductivity becomes infinity for a noninteracting system. In the lowest order of perturbation theory, we have the result
\be
\label{Kubo,0}
 \sigma^{\rm Kubo,0}_{\rm dc}=
 \frac{n e^2}{m} \frac{1}{\epsilon}
\ee
which diverges in the limit $\epsilon \to 0$.
Perturbation theory cannot be applied immediately to evaluate the dc conductivity for interacting charged particles. We discuss the use of perturbation theory for the Kubo formula in the following section \ref{Green}.\\

\noindent{\it Force-force correlation function.}
As already demonstrated in the previous section, the electrical current can be considered as a
 relevant variable to characterize the nonequilibrium state, 
when a charged particle system is affected by an electrical field.
Since the total momentum is related to the electrical current, we can select it as the relevant observable ${\rm B}_n\to \op{P}$. 
Now,  the character of Eq. (\ref{PREA15}) is changed. According  the response equation (\ref{ch6200}) we have 
\be
\label{respF}
-\langle {\rm P};\dot {\rm P} \rangle_{i \epsilon}F_n+\langle {\rm P};{\rm P} \rangle_{i \epsilon}\frac{e}{m} E=0
\ee 
so that these contributions compensate each other.
As a relevant variable, the averaged current density is determined by the response parameter $F$ which follows from the solution of the response equation (\ref{respF}).
We obtain the inverse conductivity, the resistance, as a force-force autocorrelation, see Eq. (\ref{ch6350}).
Now, perturbation theory can be applied, and in Born approximation a standard result of transport theory is obtained, the Ziman formula (\ref{ch6420}). We conclude that the use of relevant observables gives a better starting point for perturbation theory. In contrast to the Kubo formula that starts from thermal equilibrium as initial state, the correct current is already reproduced in the initial state and must not be created by the dynamical evolution.

However, despite the excellent results using the Ziman formula in solid an liquid metals where the electrons are strongly degenerate, we cannot conclude that the result (\ref{Ziman0})
 for the conductivity is already correct for low-density plasmas (non-degenerate limit if $T$ remains constant) in the lowest order of perturbation theory considered here. The prefactor $3/(4\sqrt{2 \pi})$ is wrong. If we go to the next order of interaction, divergent contributions arise. These divergences can be avoided performing a partial summation, that will also change the coefficients in Eq. (\ref{Ziman0}) which are obtained in the lowest order of the perturbation expansion. The divergent contributions can also be avoided extending the set of relevant observables $\{ {\rm B}_n\}$, see below.

Formally, it can be shown that the expression for the resistance (\ref{ch6350}) and the Kubo formula (\ref{kubo}) are consistent. We  apply partial integrations, $ \langle \dot{\op{P}} ;\dot{\op{P}} \rangle_{\ii \epsilon}
=( \dot{\op{P}} |\op{P})-\epsilon  \langle \dot{\op{P}} ;\op{P} \rangle_{\ii \epsilon}$ 
where $( \dot{\op{P}} |\op{P})= \langle [\op{P},\op{P}] \rangle=0$, and  $ \langle \dot{\op{P}} ;\op{P} \rangle_{\ii \epsilon}
=-(\op{P}|\op{P})+\epsilon  \langle \op{P} ;\op{P} \rangle_{\ii \epsilon}$ so that besides the Kubo scalar products only the momentum autocorrelation function occurs,
	\begin{equation}
	\label{sigFF}
	\sigma_{\rm dc}=\frac{n e^2}{m}\frac{mN/\beta+\langle\op{P};\dot{\op{P}}\rangle_{\ii\epsilon}}{\langle\dot{\op{P}};\dot{\op{P}}\rangle_{\ii\epsilon}}=\frac{n e^2}{m}\frac{-\langle\op{P};{\op{P}}\rangle_{\ii\epsilon}}{-(\op{P}|\op{P})+\epsilon  \langle \op{P} ;\op{P} \rangle_{\ii \epsilon}}.
	\end{equation}
 Assuming that the momentum autocorrelation function is finite, in the limit $\epsilon \to 0$ we can drop this term in the expression $-(\op{P}|\op{P})+\epsilon  \langle \op{P} ;\op{P} \rangle_{\ii \epsilon}$ so that with Eq. (\ref{ch6380}) the Kubo formula (\ref{kubo}) is recovered.\\

\noindent{\it Higher moments of the single-particle distribution function.}
Besides the electrical current, also other deviations from thermal equilibrium can occur in the stationary nonequilibrium state such as a thermal current. In general, for homogeneous systems we can consider arbitrary moments of the single-particle distribution function 
\begin{equation}
\op{P}_n= \sum_{\bs p} \hbar p_x (\beta E_p)^{n/2} {\rm a}^\dagger_{\bs p}{\rm a}^{}_{\bs p}
\end{equation}
 as set of relevant observables $\{ {\rm B}_n\}$. 
 It can be shown that with increasing number of moments the result 
\be
\label{Ziman1}
\sigma_{\rm dc}=s \frac{(k_B)^{3/2} (4 \pi \epsilon_0))^2}{m^{1/2} e^2 }\frac{1}{\Lambda( p_{\rm therm})}
\ee
is improved, as can be shown with the Kohler variational principle, see \cite{Redmer97,rr12}. The value $s=3/(4 \sqrt{2 \pi})$ obtained from the 
single moment approach is increasing to the limiting value $s=2^{5/2}/\pi^{3/2}$.
 For details see \cite{Redmer97,r90,Reinholz05}, where also other  thermoelectric effects in plasmas are considered.\\

\noindent{\it Single-particle distribution function and}
{\it the general form of the linearized Boltzmann equation.}
Kinetic equations are obtained if the occupation numbers $\op{n}_\nu$ of single-(quasi-) particle states $| \nu \rangle$ is taken as the set of relevant observables $\{\op{B}_n\}$. 
The single-particle state  $\nu$ is described by a complete set of quantum numbers,  e.g. the momentum,  the spin and the species in the case of a homogeneous multi-component plasma. 
In thermal equilibrium, the averaged occupation numbers of the quasiparticle states 
are given by the Fermi or Bose distribution function, 
$\langle {\rm n}_{\nu}\rangle_{\rm eq}=f_{\nu}^0={\rm Tr}\,\{\rho_{\rm eq} {\rm n}_{\nu}\}$.
These equilibrium occupation numbers are changed under the influence of the external field. 
We consider the deviation $\Delta {\rm n}_{\nu}={\rm n}_{\nu}-f_{\nu}^0$ as relevant observables. They describe the fluctuations of the occupation numbers. 
The response equations, which eliminate the corresponding response parameters $F_\nu(t)$, have the structure of a linear system of coupled Boltzmann equations for the quasiparticles,
see Ref. \cite{rr12} 
\begin{equation}
\label{4:65}
\frac{e}{m}{\bs E}\cdot[(\bs{{\rm P}}|{\rm n}_{\nu})+\langle \bs{{\rm P}};\dot{{\rm n}}_{\nu}\rangle_{\omega+i\epsilon}]=\sum_{\nu'}F_{\nu'}P_{\nu'\nu}\,,
\end{equation}
with
\begin{equation}
P_{\nu'\nu}=(\dot{{\rm n}}_{\nu'}|\Delta {\rm n}_{\nu})+\langle\dot{{\rm n}}_{\nu'};\dot{{\rm n}}_{\nu}\rangle_{\omega+\ii\epsilon}+i\omega\{(\Delta {\rm n}_{\nu'}|\Delta {\rm n}_{\nu})-\langle\dot{{\rm n}}_{\nu'};\Delta {\rm n}_{\nu}\rangle_{\omega+\ii\epsilon}\}\,.
\end{equation}
The response parameters  $F_\nu(t)$ are related to the averaged occupation numbers as
\begin{equation}
f_{\nu}(t)={\rm Tr}\,\{\rho(t){\rm n}_{\nu}\}=f_{\nu}^0+\beta\sum_{\nu'}F_{\nu'}(\Delta {\rm n}_{\nu'}|\Delta {\rm n}_{\nu})\,.
\end{equation}

The general form of the linear Boltzmann equation (\ref{4:65}) can be compared with the expression obtained from kinetic theory. The left-hand side can be interpreted as the drift term, where self-energy effects 
are included in the correlation function $\langle \bs{{\rm P}};\dot{{\rm n}}_{\nu}\rangle_{\omega-i\epsilon}$. In the static case $\omega = 0$, the collision operator is given by $\langle\dot{{\rm n}}_{\nu'};\dot{{\rm n}}_{\nu}\rangle_{\ii\epsilon}$.
Because the operators ${\rm n}_{\nu}$ are commuting, from the Kubo identity follows
$(\dot{{\rm n}}_{\nu'}| {\rm n}_{\nu})=(1/\hbar \beta) \langle [{\rm n}_{\nu'}, {\rm n}_{\nu}] \rangle=0$. 
More precisely, the collision operator can be expressed in terms of the correlation function of the stochastic part of fluctuations, cf. Eq. (\ref{stochforce}) below.

In the general form, the collision operator is expressed in terms of equilibrium correlation functions of fluctuations that can be evaluated by different many-body techniques. In particular, for the Lorentz model the result (\ref{Ziman1}) with $s=2^{5/2}/\pi^{3/2}$ is obtained \cite{Redmer97,r90,Reinholz05}. Furthermore, compared with KT \cite{rr12}, within LRT considered here no problems arise if the high-frequency behavior of the dielectric function (bremsstrahlung) is calculated.\\

\noindent{\it Two-particle distribution function, bound states.}
Even more information is included if we also consider the non-equilibrium two-particle distributions. As an example we mention the Debye-Onsager relaxation effect, see \cite{Roep88,r90}. Another important case is the formation of bound states. 
It seems naturally to consider the bound states as new species and to include the occupation numbers 
(more precisely, the density matrix) of the bound particle states in the set of relevant observables \cite{R81,Adams07}. It needs a long memory time to produce bound states from free states dynamically in a low-density system, because bound states cannot be formed in binary collisions, a third particle is needed to fulfill the conservation laws. The approach where bound states are considered like a new species of particles in a weakly interacting system is denoted as the chemical picture. 

The inclusion of initial correlation to improve the kinetic theory, in particular to fulfill the conservation of total energy, is an important step worked out during the last decades, 
see \cite{Mor99} where further references are given. Other approaches to include correlations in the kinetic theory are given, e.g., in Refs. \cite{ZM,MRSP}. 
Because we focus to the plasma conductivity, these more general issues are not detailed here.\\

{\it Virial expansion of the plasma conductivity}.
Based on the discussions in the previous section, we expect for the electrical conductivity of a charged particle system 
the following low-density expansion \cite{Roep88,Roepke89}
\begin{equation}
\label{virial}
 \sigma^{-1}(T,n)=A(T)\, \ln\,n+B(T)+C(T)\,n^{1/2}\ln\,n \pm \dots
\end{equation}
with 
\begin{equation}
 A(T)= -\frac{1}{2 s}\frac{e^2 m^{1/2}}{(4 \pi \epsilon_0)^2 (k_{\rm B}T)^{3/2}}\,.
\end{equation}
 We would like to stress that the 
first coefficient $A(T)$, i.e. $s=2^{5/2}/\pi^{3/2} $  is an exact result for the Lorentz plasma.
Expressions for the higher virial coefficients $B(T), C(T)$ are found in Ref. \cite{Roep88}, 
their exact values are under discussion.

Working in lowest order of perturbation theory and using only a restricted set of relevant observables ,
we obtain approximations which are not exact but may be understood as variational solutions. 
For instance, working with only one moment as in the case of the force-force correlation function,
and evaluating the correlation function in Born approximation, only the approximation $s=3/(4 \sqrt{2 \pi})$ is obtained. Improving the single-moment
Born approximation considering dynamical screening or strong collisions \cite{Roep88,Redmer97,Reinholz05},
the values of the Coulomb logarithm, in particular the values of $B(T), C(T)$, are modified. 
To get the correct value of $A(T)$, we have to consider higher terms of the perturbation expansion which are divergent. After partial summation we can expect that the correct value appears. 
As alternative which is more physical, we can extend the set of relevant observables, taking higher moments or the single-particle occupation numbers as relevant observables as discussed above.

Compared with the Lorentz plasma where the electrons interact only with the ions (fixed positions), 
more interesting is the case of a Coulomb plasma where all charged components interact. The effect of the electron-electron interaction on the dc conductivity has been discussed controversially because in the force-force correlation function no contribution appears.
The total momentum of the electron subsystem is not changed because the total momentum is conserved in $e - e$ interactions.
However, considering higher moments, the Spitzer result $s=0.591$ \cite{Spitzer53} is obtained. For discussion see \cite{Redmer97,rr12}.
Also in this case, a simple approximation is improved by summing up higher order singular terms of the perturbation expansion
which is quite complex, or working with an extended set of relevant observables. \\

{\it Hopping conductivity}.
A similar problem arises when calculating the hopping conductivity \cite{ChrisRoep85,r90}. The Hamiltonian contains the contribution of electrons which are bound 
in localized states at (disordered) ion positions ${\bs R}_i$ as well as transfer matrix elements describing tunneling (hopping) of the
localized electrons. A simple evaluation of the Kubo formula (\ref{kubo}) gives a finite result, which is, however, only an approximation.
The correct result for the hopping conductivity is obtained if the local occupation numbers
$ {\rm n}_i$ at the ion positions ${\bs R}_i$ are taken as relevant observables. This problem has been discussed controversially in the literature, see \cite{ChrisRoep85,R81},  but the situation becomes clear considering 
the Zubarev approach allowing for an extended set of relevant observables, the occupation numbers of the localized states. The corresponding response parameters $\lambda_i$ may be considered as local chemical potentials.

\section{Green functions approach for the Kubo formula}
\label{Green}

{\it Response functions and thermodynamic Green functions}.
We investigate the question whether we can sum up the perturbation expansion to obtain correct results for the conductivity,
even if we start from an expression for the conductivity which is obtained from a reduced set of relevant observables 
$\{ {\rm B}_n\}$,
as mentioned above in Sec. \ref{Operatorset}. 
Starting from a coarse description of the nonequilibrium situation by the corresponding relevant statistical operator 
$\rho_{\rm rel}(t)$, the missing correlations must be produced dynamically. 
This means that we have to consider higher orders of the perturbation expansion of the time evolution operator. 
In the present section, we demonstrate this considering a simple case, the conductivity $\sigma(\omega)$. 
We show that higher order perturbation theory and partial summations are necessary to get an acceptable result 
(i.e. also an approximation), the force-force correlation result in Born approximation presented in Sec. \ref{conductivity}, 
even if we start from the simplest case, the empty set of relevant observables $\{ {\rm B}_n\}$, i.e. the Kubo formula.
Within a quantum statistical approach, we use the method of thermodynamic Green functions.
As example, below we consider the Lorentz model. 

We start from the Kubo formula (\ref{kubo}) (the factor 1/3 appears owing to  the vector representation and isotropy)
\begin{equation}
\label{Kubo0}
 \sigma^{\text{Kubo}}(\omega)= \frac{e^2 \beta}{3m^2 \Omega} \langle {\bf P}; {\bf P}  \rangle_{\omega+i\epsilon}^{\text{irred}}
\end{equation}
with $
{\bf P} = \sum_{\bs p} \hbar {\bs p}\, {\rm a}_{\bs p}^+{\rm a}_{\bs p}^{} $, Eq. (\ref{ch6371}),
so that
\begin{equation}
\label{GFKubo}
 \sigma^{\text{Kubo}}(\omega)= \frac{e^2 \beta \hbar^2}{3m^2 \Omega} 
 \sum_{\bs p,\bs p'} {\bs p} \cdot {\bs p}' \langle {\rm n}_{\bs p}; {\rm n}_{\bs p'}  \rangle_{\omega+i\epsilon}^{\text{irred}}.
\end{equation}

We have to calculate the Laplace transform of the correlation function:
\begin{equation}
 \langle {\rm a}_{\bs p}^+{\rm a}_{\bs p}^{}; {\rm a}_{\bs p'}^+{\rm a}_{\bs p'}^{} \rangle_z = \int \limits_0^{\infty} \dif t \,\e^{\ii zt} 
 \int \limits_0^1 \dif \lambda \,\mbox{Tr}\{\e^{\frac{\ii}{\hbar}\Hop_{\rm S}(t-i\hbar\beta\lambda)}{\rm a}_{\bs p}^+{\rm a}_{\bs p}^{}
\, \e^{-\frac{\ii}{\hbar}\Hop_{\rm S}(t-i\hbar\beta\lambda)}{\rm a}_{\bs p'}^+{\rm a}_{\bs p'}^{}\rho_{\rm eq}\}.
\end{equation}
The time dependence as well as the equilibrium statistical operator contain the system Hamiltonian 
$\Hop_{\rm S}=\Hop_0+\Hop_{\rm int} $ (\ref{ch6370}). 
For the case of a charged particle system, the Coulomb interaction between the plasma components can be taken as the 
interaction part. 
For the Lorentz model, only the electron-ion interaction (\ref{Debpot}) is considered.

To perform a systematic evaluation of the correlation functions arising in linear response theory, we use the method of thermodynamic Green functions \cite{ZubarevGF}. The thermodynamic Green 
function of  operators ${\rm A},{\rm B}$ is defined as
\begin{equation}
G_{A,B}(\ii z_\lambda)=\int_0^\beta \dif \tau \,  \e^{\ii z_\lambda \tau} \langle {\rm T}\{ {\rm A}(\tau) {\rm B} \} \rangle
\end{equation}
where we introduce the Heisenberg-like dependence on the parameter $\tau$ according to
${\rm A}(\tau)= e^{\tau (\Hop_0-\sum \mu_c {\rm N}_c)}{\rm A}e^{-\tau (\Hop_0-\sum \mu_c {\rm N}_c))}$;
the ${\rm T}\{\dots \}$-product denotes the ordering of operators
with growing parameter values $\tau$
from right to left. The Fourier transform is defined at the bosonic Matsubara frequencies $z_\lambda=\pi \lambda/\beta$,
$\lambda = 0,\,\, \pm 2, \dots$ are the even numbers.
Analytical continuation from $z_\lambda$ into the whole complex $z$-plane
gives the  spectral function ${\rm Im}\,G_{A,B}(\omega+\ii \epsilon)$ at $z=\omega+\ii \epsilon$.

We use the following relation
\begin{equation}
\label{KuboVI}
 \langle {\rm A};{\rm B} \rangle_{z}=\frac{\hbar}{\beta}\int \frac{\dif \omega'}{\ii \pi}\,\frac{1}{z-\omega'} \,\frac{1}{\omega'} 
{\rm Im}\,G_{A,B}(\omega'+\ii \epsilon).
\end{equation}
A similar relation can be derived also for the Kubo scalar product.

For the Kubo formula (\ref{Kubo0}) we have
\begin{equation}
\label{Kubo1}
 \sigma^{\text{Kubo}}(\omega)= \frac{e^2 \beta}{3m^2 \Omega} \frac{\hbar}{\beta}\int \frac{\dif \omega'}{\ii \pi }\,\frac{1}{z-\omega'} \,\frac{1}{\omega'} 
{\rm Im}\,G_{\bs P,\bs P}(0,\omega'+\ii \epsilon)
\end{equation}
with
\begin{eqnarray}
\label{GPP}
G_{\bs P,\bs P}({\bs Q},\ii Z_\lambda)&=&\hbar^2\sum_{{\bs p}{\bs p}'}{\bs p}\cdot{\bs p}'\sum_{z_\nu,z_\nu'}
\Pi({\bs p},\ii z_\nu,{\bs Q},\ii Z_\lambda,{\bs p}',\ii z_\nu').
\end{eqnarray}

Using Feynman diagrams, the polarization function $\Pi({\bs p},\ii z_\nu,{\bs Q},\ii Z_\lambda,{\bs p}',\ii z_\nu')$ 
is the sum of all irreducible diagrams with a left free vertex, incoming propagator ${\bs p},\ii z_\nu$, 
outgoing propagator ${\bs p}+{\bs Q},\ii z_\nu+\ii Z_\lambda$, and a right free vertex,
incoming propagator ${\bs p}'+{\bs Q},\ii z_\nu'+\ii Z_\lambda$, 
outgoing propagator ${\bs p}',\ii z_\nu'$. 
The polarization function is related to the dielectric function $\varepsilon({\bs Q},\ii Z_\lambda)$ at wave vector 
$ {\bs Q} $ and bosonic Matsubara frequency $ \ii Z_\lambda $. The simplest diagram is the well-known RPA loop.

The analytical continuation $\ii Z_\lambda \to z$ and taking $z=\omega +\ii \epsilon$ as well as the limit ${\bs Q} \to 0$ yields 
the dynamical conductivity in the long-wavelength limit. The static (dc) conductivity follows for ${\bs Q}=0$
and $z \to \ii \epsilon$.\\

{\it Zeroth order with respect to the interaction}.
In lowest order of the interaction, the polarization function is given by the random phase approximation (RPA),
\begin{align}
\label{RPA}
\sum_{z_\nu,z_\nu'}
\Pi^{\rm RPA}({\bs p},\ii z_\nu,{\bs Q},\ii Z_\lambda,{\bs p}',\ii z_\nu')
&= \frac{f(E_{ p})-f(E_{\bs p+\bs Q})}{\ii Z_\lambda +E_{ p}- E_{\bs p+\bs Q}}\delta_{\bs p,\bs p'}\,.
\end{align}
Analytical continuation in the complex $z$ plane and approaching the real axis from above 
($\ii Z_\lambda \rightarrow z=\omega+\ii \epsilon$) gives a jump of the imaginary part if
we cross the real axis. This determines the 
spectral density
\begin{align}
I_{n_{\bs p},n_{\bs p'}}(\omega, Q) 
&=2\pi f(E_p)[1-f(E_p)] \delta_{\bs p,\bs p'}\delta(\omega+E_p-E_{\bs p+\bs Q})\,.
\end{align}
After Fourier transformation we have for the Laplace transform 
\begin{align}
\langle {\rm a}_{\bs p}^+{\rm a}_{\bs p}^{}; {\rm a}_{\bs p'}^+{\rm a}_{\bs p'}^{}  \rangle_{i\epsilon} 
=\frac{1}{\epsilon}f(E_p)[1-f(E_p)]\delta_{\bs p,\bs p'}
\end{align}
and finally
\begin{equation}
 \sigma^{\text{Kubo},0}(\omega = 0)= \frac{e^2 \beta \hbar^2}{3m^2 \Omega} 
 \sum_{\bs p,\bs p'} {\bs p} \cdot {\bs p}' \frac{1}{\epsilon}f(E_p)[1-f(E_p)]\delta_{\bs p,\bs p'}
 = \frac{e^2 \beta \hbar^2}{3m^2 \Omega} 
\int \frac{\dif^3p \,\Omega}{(2 \pi)^3} p^2 \frac{1}{\epsilon}f(E_p)[1-f(E_p)]\,.
\end{equation}
Using integration by parts, the integral can be performed with the result (\ref{Kubo,0}) $ \sigma^{\rm Kubo,0}_{\rm dc}= n e^2/(m \epsilon)$.
Obviously, as already mentioned, the lowest order of perturbation theory is diverging when $\epsilon \to 0$. In this approximation, the electrical current is conserved, and the the correlation function is not time dependent.\\

{\it Dressed propagators}.
Considering higher orders of perturbation theory, we replace the free propagators 
by dressed propagators determined by the self-energy,
\begin{equation}
 G({\bs p},\ii z_\nu)= \int \frac{\dif \omega}{2 \pi} \frac{1}{\ii z_\nu-\omega}A({\bs p},\omega)= \int \frac{\dif \omega}{2 \pi}\, \frac{1}{\ii z_\nu-\omega} \,\frac{{\rm Im} \Sigma({\bs p},\omega)}{[\omega 
 -E_p-{\rm Re}\Sigma({\bs p},\omega)]^2 +[{\rm Im} \Sigma({\bs p},\omega)]^2}\,.
 \end{equation}
For this, we need the expression for the self-energy $\Sigma({\bs p},iz_\nu)$.

In particular, we use again the Born approximation where in lowest order of density
\begin{eqnarray}
&&\Sigma({\bs p},\ii z_\nu)=\frac{1}{\beta^2}\sum_{\bs q,\bs k,\Omega_\mu,z_\nu'}V^2(q)\frac{1}{\ii z_\nu'-
E^{}_{\bs p+\bs q}}\,\frac{1}{\ii  \Omega_\mu-\ii z_\nu'-E^{\rm ion}_{\bs k-\bs q}}\,\frac{1}{\ii \Omega_\mu-\ii z_\nu-E^{\rm ion}_{k}}\nonumber\\&&
=\sum_{\bs q,\bs k}V^2(q)\frac{f(E^{\rm ion}_k)}{\ii  z_\nu+E^{\rm ion}_{k}-E_{\bs p+\bs q}-E^{\rm ion}_{\bs k-\bs q}}+{\cal O}(n)\,
\end{eqnarray}
(begin with $\sum_{z_\nu'}$ and neglect $f(E_{\bs p+\bs q}),\,\,f(E^{\rm ion}_{\bs k-\bs q}) \ll 1$). In the adiabatic limit where the collisions of electrons with the ions are quasi elastic, 
the contribution $E^{\rm ion}_{k}-E^{\rm ion}_{\bs k-\bs q}$ can be dropped, and we have
\begin{equation}
\label{tauV}
\Sigma({\bs p},E_p)=\sum_{\bs q,\bs k}V^2(q)\frac{f(E^{\rm ion}_k)}{E_{p}-E_{\bs p+\bs q}}
\end{equation}
so that
\begin{equation}
\label{Imtau}
{\rm Im} \Sigma({\bs p},E_p)=\frac{\hbar}{2 \tau_p}=
N_{\rm ion}\sum_{\bs q}V^2(q)\, \pi \delta(E_{p}-E_{\bs p+\bs q})
\end{equation}
that explicitly expresses the energy conservation during collisions with ions in the adiabatic limit.

We approximate the polarization function by the product of two full single-particle Green functions, see Sec. \ref{PiGG}.
We have for $Z_\lambda \to \omega + \ii \epsilon$
\begin{equation}
\label{ImPi}
{\rm Im} \sum_{z_\nu,z_\nu'}\Pi^{GG}({\bs p},\ii z_\nu,{\bs Q},\omega,{\bs p}',\ii z_\nu')=
\frac{1/\tau_p+1/\tau_{\bs p+\bs Q}}{(\omega +E_{\bs p+\bs Q} -E_{p})^2+(1/\tau_p+1/\tau_{\bs p+\bs Q})^2}
[f(E_{p})-f(E_{\bs p+\bs Q})] \delta_{{\bs p},{\bs p}'}.
\end{equation}
The spectral function 
and its Laplace transform lead to the result
\begin{eqnarray}
&&\langle {\rm n}_{\bs p}; {\rm n}_{\bs p'} \rangle_{\ii \epsilon} 
\delta_{{\bs p},{\bs p}'}\nonumber \\ &&
= \frac{2/\tau_p}{(E_{\bs p+\bs Q} -E_{p})^2+4/\tau_p^2}f(\epsilon_{p})[1-f(E_{\bs p+\bs Q})] \delta_{{\bs p},{\bs p}'}=
\frac{1}{2} \tau_p f(E_{p})[1-f(E_{\bs p+\bs Q})] \delta_{{\bs p},{\bs p}'}
\end{eqnarray}
so that with Eq. (\ref{GFKubo})
the contribution
\begin{equation}
 \sigma^{\text{Kubo},1}(0)= \frac{e^2 \beta \hbar^2}{3m^2 \Omega} 
 \sum_{\bs p} p^2  \frac{1}{2} \tau_p f(E_{p})[1-f(E_{p})]= \frac{n e^2}{m^2 }{\bar \tau}
\end{equation}
follows from Eq. (\ref{Kubo1}). 
We introduced the average total cross section ${\bar \tau}$
given by $\tau_p$, Eq. (\ref{Imtau}), at an appropriate value of $p$. 
This  result for $\sigma_{\rm dc}$ is finite, but incorrect. Instead of the total cross section ${\bar \tau}$, 
the transport cross section should appear. This shows that the evaluation within perturbation expansions should be performed with care. Results are obtained within a certain order of 
the perturbation which are not exact within the considered order of the perturbative expansion, but only some approximations. We have to take all relevant contributions, see the following subsection. Such so-called conserving approximations are known from the general theory \cite{KB} of thermodynamic Green functions.\\

{\it Vertex contribution}.
To be consistent, we have to consider further diagrams which are of the same order as the self-energy terms, the vertex corrections.
To sum up such contributions, we consider the Bethe-Salpeter equation (BSE)
\begin{eqnarray}
\label{BSE}
 && \Pi({\bs p},\ii z_\nu,{\bs Q},\ii Z_\lambda,{\bs p}',\ii z_\nu')=\int \frac{\dif \omega_1}{2 \pi} \int \frac{\dif \omega_2}{2 \pi}  
 A({\bs p},\omega_1) A({\bs p-Q},\omega_2)\,
 \frac{1}{\ii z_\nu-\omega_1} \,\frac{1}{\ii  z_\nu-\ii Z_\lambda-\omega_2 } \nonumber\\ 
 && \times
 \left\{\delta_{\bs p,\bs p'}\delta_{z_\nu,z_\nu'}+\sum_{\bs p_1,z_1}\Gamma({\bs p},\ii z_\nu,{\bs Q},\ii Z_\lambda,{\bs p}_1,\ii  z_1)
\,\Pi({\bs p}_1,\ii z_1,{\bs Q},\ii Z_\lambda,{\bs p}',\ii z_\nu')\right\}.
\end{eqnarray}
In addition to the product of full single particle propagators $G({\bs p},\ii z_\nu)$, see Eq. (\ref{ImPi}), 
the effective interaction kernel $\Gamma({\bs p},\ii z_\nu,{\bs Q},\ii Z_\lambda,{\bs p}_1,\ii  z_1)$ is introduced which 
can be represented by the corresponding irreducible diagrams.
In contrast to the Dyson equation or the screening equation where an algebraic solution can be given, the BSE (\ref{BSE}) is an integral equation because in $\Pi$ the variables ${\bs p},\ii z_\nu$ are changed to  ${\bs p}_1,\ii z_1$ that have to be integrated 
after multiplication with $\Gamma$. To solve it we make some simplifications concerning the dependence on ${\bs p}_1,\ii z_1$ which are given in the Appendix \ref{app:vertex}.

As a result we find with Eq. (\ref{transp})
\begin{equation}
\label{Kubo5}
 \sigma^{\text{Kubo,2}}(0)= \frac{n e^2}{m}\bar{\tau}^{\rm transp}\,,
\end{equation}
the total cross section has been replaced by the transport cross section.

We conclude that the perturbative approach is rather cumbersome, but it gives some insight 
how the Zubarev NSO approach works. 
The result which is immediately obtained from the force-force correlation function is not easily reproduced.
It is also clear that this result is not correct because the Coulomb logarithm has not the correct prefactor.
If we go to higher orders of perturbation theory, more effort is necessary, and possibly we can find the 
correct prefactor. It is more simple to work with the occupation numbers as relevant observables which
gives the kinetic equations, and the solution of the Lorentz plasma conductivity is found 
using the relaxation time ansatz.

The Lorentz plasma model is quite simple so that the perturbation expansion can be summed up. Solving the Schr{\"o}dinger equation for the electron moving in the potential of a fixed ion configuration, we can 
replace the RPA expression (\ref{RPA}) for the polarization function by the atomic loop 
using the full electron-ion two-particle propagator \cite{RD}. 
This corresponds to the Kubo-Greenwood formula to be discussed in the following section.

A more realistic approach to the conductivity of a plasma should include $e - e$ collisions
\cite{Spitzer53}. For this,
higher orders of perturbation theory have to be considered what is beyond the present approach of this section.
Note that also in using the Kubo-Greenwood formula as done in the next section, the inclusion of  $e - e$ collisions
is not solved yet, see \cite{rrrr15}. In contrast, within generalized linear response theory 
the account of  $e - e$ collisions is no problem.\\

{\it The Kubo-Greenwood approach}.
We can avoid perturbation expansions applying numerical solutions of the many-particle system.
This is done, for instance, using the (classical) molecular dynamics (MD) simulations. In solids,
a quantum treatment is obtained from the density functional theory (DFT). 
Starting point for the calculation of the conductivity in the DFT-MD method is the Kubo 
formula (\ref{KuboVI}). The equilibrium statistical operator $ \rho_{\rm eq}$ contains the Kohn-Sham (KS) 
Hamilton operator $ {\rm H}_{\rm KS}$, see \cite{rrrr15}. As a particular example, we can also consider the Lorentz plasma with
given ion configuration and solving the Schr{\"o}dinger equation for the Coulomb electron-ion potential. 
We only briefly discuss this approach to calculate conductivity 
in complex systems which became quite popular nowadays. For further references see \cite{rrrr15}.

Within the DFT-MD method, the system of noninteracting electrons in the potential of ions at given positions
(configuration) ${\bs R}_i$ is treated solving the effective single-particle Schr{\"o}dinger equation
\begin{equation}
{\rm H}_{\rm KS} |{\bs k} \nu \rangle =E_{{\bs k}\nu} |{\bs k}  \nu \rangle
\end{equation}
numerically. A finite number of electrons and ions is considered, 
and periodic boundary conditions are implemented. Discrete wave numbers $\bs k$ 
are given by the periodic boundary conditions, and a bound state level splits into 
subbands $\nu$ according to the number of ions within the periodic cell $\Omega_c$.
The time-dependence of the operators within the Heisenberg 
picture in the momentum autocorrelation function is treated as
\begin{equation}
{  {\bf P}}(t-\ii \hbar \tau) =\e^{\frac{\ii}{\hbar}(t-\ii \hbar \tau) {\rm H}_{\rm KS}} {  {\bf P}}
\e^{-\frac{\ii}{\hbar}(t-\ii \hbar \tau)  {\rm H}_{\rm KS}}.
 \end{equation}
The momentum operator 
reads in second quantization with respect to this basis
$
{ {\bf P}}= \sum_{{\bs k,k'},\nu,\nu'} \langle {\bs k} \nu| { {\bf p}}| {\bs k'} \nu' \rangle
 {\rm a}^\dagger_{{\bs k} \nu}   {\rm a}^{}_{{\bs k}' \nu'}
$. 
The matrix elements are given by
\begin{equation} 
\langle {\bs k} \nu| { {\bf p}}| {\bs k'} \nu' \rangle=\delta_{\bs k ,\bs k'}\left[\hbar {\bs k} \,\delta_{\nu, \nu'}
+\frac{1}{\Omega_c} \int_{\Omega_c} \dif^3 {\bs r}\, u^*_{ {\bs k} \nu}( {\bs r})  \frac{\hbar}{\ii}
\frac{\partial }{\partial { {\bs r}}} u_{ {\bs k} \nu'}( {\bs r}) \right].
\end{equation} 
In this representation, the time-dependence of the momentum operator is immediately given.
The average with the equilibrium statistical operator is 
evaluated using Wick's theorem. 
From the Kubo formula (\ref{KuboVI}), we find for the real part of 
the optical conductivity tensor
\begin{eqnarray}\label{KG1}
 {\rm Re}\, 
 \sigma^{\rm KG}_{\alpha \beta}(\omega)&=&\frac{2 \pi e^2}{3 \Omega_{c} m^2 \omega} \sum_{{\bs k}\nu\nu'}
 \langle {\bs k} \nu| { {\bf p}}_\alpha| {\bs k} \nu' \rangle \cdot 
\langle {\bs k} \nu'| { {\bf p}}_\beta| {\bs k} \nu \rangle
 (f_{ {\bs k} \nu}-f_{ {\bs k} \nu'}) \delta_\epsilon (E_{ {\bs k} \nu}-E_{ {\bs k} \nu'}-\hbar \omega) \,.
\end{eqnarray}

The numerical evaluation of the dc conductivity $\lim_{\omega \to 0}\sigma^{\rm KG}_{\alpha \beta}(\omega)$
is intricate because a value $0/0$ appears.
Therefore, a broadened $\delta$ function
\begin{equation}\label{delta}
 \delta_\epsilon(x) = \frac{1}{\pi}\frac{\epsilon}{x^2+\epsilon^2} 
\end{equation}
is introduced which  makes a smooth transition in the static case  ($\omega \to 0$ ).  
For the application of the Kubo-Greenwood formula given as Eq.~(\ref{KG1}),
because of the finite simulation volume $\Omega_c$
and resulting discrete eigenvalues, the $ \delta$-function must
be broadened. For instance, a Gaussian broadening of the
$\delta$-function can be used that is as small as feasible without recovering the
local oscillations in the optical conductivity resulting from the discrete band structure
(see citation in Ref. \cite{rrrr15}).

The finite width of the $\delta_\epsilon(x)$ function can be interpreted as 
an additional damping to overcome the 
level spacing due to the finite volume with periodic boundary conditions. 
The limit $\epsilon \rightarrow 0$ can be taken only in the final expressions, summing up all orders 
of perturbation expansion. Expanding with respect to the electron-ion interaction $ {\rm V}$, 
the van Hove limit ($ {\rm V}^2/\epsilon \to 0$) has to be taken. 
Then, for finite $\epsilon$ a perturbation expansion of (\ref{KG1}) can be performed.

With the perturbation expansion (no formation of subbands $\nu$)
\begin{equation}
\langle {\bs k}_1|{ {\bf p}}|{\bs k}_2 \rangle = \hbar {\bs k}_1 \delta_{{\bs k}_1,{\bs k}_2} +
\frac{\langle {\bs k}_1| {\rm V}|{\bs k}_2 \rangle}{E_{{\bs k}_1}-E_{{\bs k}_2}} (\hbar {\bs k}_1- \hbar {\bs k}_2)
\end{equation}
we have with ${\bs k}_2={\bs k}_1+{\bs q}$ and $\langle {\bs k}_1| {\rm V}|{\bs k}_2 \rangle=V_q$
\begin{eqnarray} \label{xxx}
{\rm Re}\, \sigma^{\rm KG}(0) &=& \frac{\pi e^2 \hbar}{3 m^2 \Omega}\sum_{\bs k,\bs q} \frac{\partial f(E_k)}{\partial E_k}
\left({\bs k}\, \delta_{{\bs q},0}+\frac{V_q}{E_k-E_{\bs k+\bs q}}{\bs q}+\dots\right)^2  \frac{\epsilon}{\epsilon^2+
(E_k-E_{\bs k+\bs q})^2}.
\end{eqnarray}
Considering the screened interaction with uncorrelated singly charged ions in the nondegenerate case, 
$V_q^2 = N_{\rm ion} e^4/[\varepsilon_0 \Omega (q^2 +\kappa^2_{\rm D})]^2\, , \, \kappa^2_{\rm D}=\beta n e^2 /\varepsilon_0 $, Eq. (\ref{xxx}) leads to
\begin{eqnarray}
{\rm Re}\, \sigma^{\rm KG}(0) &=& \frac{\pi e^2 \hbar \beta}{3 m^2 }\int \frac{\dif^3 k}{(2 \pi)^3} f(E_k)
\left(k^2 \frac{1}{\epsilon}+\int\frac{\dif^3 q}{(2 \pi)^3}\frac{n_{\rm ion} Z^2 e^4}{[\varepsilon_0 (q^2 +\kappa^2)]^2 
(E_k-E_{\bs k+\bs q})^2}
q^2 \frac{\epsilon}{\epsilon^2+(E_k-E_{\bs k+\bs q})^2}+\dots\right)\nonumber \\
&=&\frac{\pi e^2 \hbar \beta}{3 m^2 }\int \frac{\dif^3 k}{(2 \pi)^3} f(E_k)
k^2 \tau^{\rm KG}(k)+\dots
\end{eqnarray}
with
\begin{eqnarray}
\tau^{\rm KG}(k)&=&\frac{1}{\epsilon}+ \frac{1}{\epsilon^2}\frac{1}{k^3}\frac{n_{\rm ion} e^4 m \pi}{\varepsilon^2_0 \hbar^2 }\int_0^{2 k}\frac{\dif q}{(2 \pi)^2}
\frac{q^3}{(q^2 +\kappa^2_{\rm D})^2 } +{\cal O}\left(\frac{e^8}{\epsilon^3}\right)\,,
\end{eqnarray}
see Ref. \cite{rrrr15}. 

In principle, one has to sum the leading divergent terms $\propto (1/\epsilon) \left(e^4/\epsilon\right)^n$. We give here only the first contributions,
\begin{equation}
\frac{1}{\epsilon}+ \frac{1}{\epsilon^2}A+\dots = \frac{1}{\epsilon}\left[1+ \frac{1}{\epsilon}A+\dots\right] = \frac{1}{\epsilon}\frac{1}{1- \frac{1}{\epsilon}A+\dots}.
\end{equation}
Now the limit $\epsilon \to 0$ can be performed with the result $-1/A$.

For comparison, see \cite{rr12}, with the golden rule for the transition rates and the structure factor 
$S(q)\approx 1$ so that $ |V_{\textrm{ei}}(q)|^2\approx V_{q}^{2} $, 
the energy dependent relaxation time can be calculated
 \begin{eqnarray} \label{reltime1}
  \frac{1}{\tau_k} &=&
 -\frac{2 \pi}{\hbar}\sum_{\bs q} V^2(q)\, \delta(E_k-E_{\bs k+\bs q})\frac{\bs E \cdot \bs q }{\bs E \cdot \bs k }.
 \end{eqnarray}
 The $\bs q$ integral in \rf{reltime1} can be performed using spherical coordinates where $\bs k$ is in $z$ direction, $\bs E$ in the $x-z$ plane.
It is convergent only in the case of a  screened Coulomb potential. Using the 
statically screened Debye potential 
$V_{q}=e^2/\{\varepsilon_0 \Omega (q^2+\kappa^2_{\rm D})\} $, 
we find the energy ($k$) dependent collision frequency
\begin{equation} \label{coulomblog}
\nu_k=\tau_k^{-1}=n \frac{ e^4}{4 \pi \varepsilon_0^2} \frac{m}{\hbar^3 k^3}\left( \ln \sqrt{1+b}-\frac{1}{2} \frac{b}{1+b}\right),
\end{equation}
with $b= 4 k^2 /\kappa^2_{\rm D}$. 
The static conductivity is determined as
\begin{eqnarray} \label{jel}
\sigma_{\textrm{dc}}^{\rm Lorentz} &=& 
\frac{e^2 \hbar^2}{m^2}\beta \frac{1}{ \Omega_0} \sum_{\bs k} \, k_E^2 \, \tau_k \,
f(E_k)\left[1-f(E_k)\right] 
= \varepsilon_0 \omega^2_{\textrm{pl}} \tau^{\rm Lorentz}=\frac{e^2 n_{\rm e}}{m\, \nu^{\rm Lorentz}}\,.
\end{eqnarray}
We introduce the average relaxation time $\tau^{\rm Lorentz}$ 
and the static collision frequency $\nu^{\rm Lorentz}=1/\tau^{\rm Lorentz}$.
The approach can also be applied for a pseudo-potential describing the $e-i$ interaction 
and an ion structure factor describing the ion configuration.
The exact solution for the dc conductivity Lorentz model can be found if using the relaxation time ansatz. It corresponds to the Brooks-Herring result (see Ref. \cite{Roep88})
where the semiconductor conductivity for the screened electron-hole interaction is considered. It is not clear yet whether the electron-electron interaction is already included in the KS Hamiltonian, or whether additional 
electron-electron scattering has to be taken into account. The KS Hamiltonian contains $e - e$ interaction only as mean field.

The Kubo-Greenwood approach can also be found from the cluster decomposition of the polarization function \cite{RD}.
Using the relation 
\begin{equation}
\sigma(\bs q,\omega) =\frac{\ii \omega}{q^2} \Pi(\bs q,\omega)
\end{equation}
$\Pi(\bs q,\omega)$ is calculated for the full solution of the electrons interacting with the 
entire ion system instead of a single ion. The corresponding single-electron states are a basis to evaluate $\Pi$.

We have seen that some finite source term $\epsilon$ is necessary to get converging results for the static conductivity.
Even avoiding the perturbation expansion, the source term in the Liouville-von Neumann equation 
is necessary to obtain a finite value for $\sigma(\omega \to 0)$. 
Basically, we are considering a single electron moving in a mean-field potential produced by the ions at fixed positions as well as by the other electrons. Correlations, in particular collisions, between the electrons are neglected. It is not clear whether the exact solution of the single-electron problem can describe a dissipative behavior, without any additional assumptions. This is a problem in any simulation of the solution of the many-particle 
dynamics.

\section{Outlook}\label{Outlook}


{\it Problems with the limit $\epsilon \to 0$}.
Within many-particle theory, improved expressions of the collision integral that determines the 
resistance can be derived, taking into account strong collisions, dynamical screening,
degeneration and structure factor effects. This would allow to calculate the coefficients of the virial expansion of $\sigma_{\rm dc}^{-1}$
(\ref{virial}), see Ref. \cite{Roep88}. The quantum statistical approach to evaluate the equilibrium 
correlation functions has been worked out for various applications.

It is not clear whether the rigorous evaluation of the correlation functions will give non-trivial results for the 
conductivity. For instance, arguments can be given that the exact evaluation of the force-force correlation function 
leads to a vanishing result.

Making use of the relations (\ref{ch6240}) between the correlation functions 
we can write the resistance \cite{Kalashnikov}
	\be\begin{array}{rcl}\displaystyle
	R&=&\displaystyle\frac{m^2\Omega}{e^2\beta(\op{P}|\op{P})}\,\frac{\langle\dot{\op{P}};\op{P}\rangle_{\ii\epsilon}}{\langle \op{P};\op{P}\rangle_{\ii\epsilon}}
	\\
	\
	\\
	&=&\displaystyle\frac{m^2\Omega}{e^2\beta(\op{P}|\op{P})}\,\frac{1}{(\op{P}|\op{P})}\left\langle\left\{\dot{\op{P}}-\frac{\langle\dot{\op{P}};\op{P}\rangle_{\ii\epsilon}}{\langle \op{P};\op{P}\rangle_{\ii\epsilon}}\op{P}\right\};\left\{\dot{\op{P}}-\frac{\langle\dot{\op{P}};\op{P}\rangle_{\ii\epsilon}}{\langle \op{P};\op{P}\rangle_{\ii\epsilon}}\op{P}\right\}\right\rangle_{\ii\epsilon}.
	\end{array}
\label{stochforce}
	\ee
For the proof use Eqs. (\ref{ch6240}). This expression contains the stochastic forces 
\be
{\rm F}_{\rm st}=\dot{\op{P}}-\frac{\langle\dot{\op{P}};\op{P}\rangle_{\ii\epsilon}}{\langle \op{P};\op{P}
\rangle_{\ii\epsilon}} \op{P}
\ee
 in analogy to the corresponding
term in the Langevin equation. 
The equivalence between
the resistance $R$ as an inverse transport coefficient, i.e.,   as a 
quantity that  expresses the dissipation
of energy in the system, on one hand, and the correlation functions  on the other hand
that give information about the fluctuations of the stochastic forces 
in equilibrium, is the so-called \emph{second 
Fluctuation-Dissipation-Theorem}. \\

{\it The memory-function approach}.
Closely related to the calculation of the inverse conductivity $R=1/\sigma$ using the force auto-correlation function is the memory-function approach according to Mori \cite{Mori}. The Kubo formula (\ref{kubo}) for the frequency dependent conductivity
\begin{equation}
 \sigma^{\rm Kubo}(\omega)=
\frac{e^2 \beta}{m^2 \Omega} \langle \op{P} ;\op{P} \rangle_{\omega+ \ii \epsilon}^{\rm irred}
\end{equation}
(note that the irreducible part has to be taken because the conductivity is defined with respect to the internal electric field 
${\bs E}^{\rm int}={\bs E}^{\rm external}/\varepsilon$ with the dielectric function $\varepsilon(\omega)$, we will not do so any further  here) is rewritten as
\begin{equation}
\label{sigKubo}
 \sigma^{\rm Kubo}(\omega)=
\ii \frac{n e^2 }{m} \frac{1}{\omega+M(\omega)}.
\end{equation}
The memory function $M(\omega)$ is given by  the "proper" part of the force-force correlation function
\begin{equation}
M(\omega)=\ii \langle\dot{\op{P}};\dot{\op{P}}\rangle^{\rm proper}_{\omega+ \ii \epsilon} \frac{\beta}{mN}=M'(\omega)+\ii M''(\omega)
\end{equation}
with  $M'(\omega)=-M'(-\omega)$ for the real part and 
 $M''(\omega)=-M''(-\omega)$ for the imaginary part, for details see Ref. \cite{Plakida}. The definition of the memory function as the  "proper" part of the force-force correlation function correspond to the introduction of the projected Liouville superoperator in the Mori approach \cite{Mori}.
Compared with Eq. (\ref{stochforce}), the projection in the time evolution operator is not easy to handle if higher orders of perturbation theory are considered.
The dynamical conductivity follows as
\begin{equation}
\label{sigMem}
 \sigma^{\rm Kubo}(\omega)=
\frac{ne^2}{m} \frac{1}{M''(\omega)-\ii(M'(\omega)+\omega)}.
\end{equation}
For $\omega =0$ we obtain the resistivity $R=M''(0)m/(ne^2)$.
Comparing with the linear response approach given in Sec. \ref{Operatorset} where we obtained for the optical conductivity
	\begin{equation}
	\label{sigFFom}
	\sigma(\omega)=\frac{n e^2}{m}\frac{1+\frac{\beta}{mN}\langle\dot{\op{P}};\op{P}\rangle_{\omega+
	\ii\epsilon}}{\frac{\beta}{mN}\langle\dot{\op{P}};\dot{\op{P}}\rangle_{\omega+\ii\epsilon}-
	\ii \omega \frac{\beta}{mN} \langle\dot{\op{P}};\op{P}\rangle_{\omega+\ii\epsilon}-\ii \omega},
	\end{equation}
see Eq. (\ref{sigFF}), the correlation functions $ \langle\dot{\op{P}};\op{P}\rangle_{\omega+\ii\epsilon}$ do not occur 
in Eqs. (\ref{sigKubo}), (\ref{sigMem}), because of the projection operator technique.
The memory-function approach has been applied to calculate the optical and dc conductivity of different systems, see Ref. \cite{Plakida} and further references given there.\\


{\it Heat production and entropy}.
Electrical conductivity describes a non-equilibrium process. On a macroscopic level, mechanical energy, represented by the electrical field, is transformed to heat. On the microscopic level, ordered motion of the electrons imposed by the external field is dispersed to disordered motion because of collisions of electrons with ions. As a consequence, an electrical current representing the collective 
motion of electrons is attenuated by collisions, what is compensated by the action of the external field. Of course, one has to consider an ensemble to obtain the average damping of the current. Linear response theory solves this problem, considering a fluctuation in equilibrium which is characterized by a current, and considers the current-current correlation function to calculate the damping rate of a small fluctuation in 
a system in thermodynamic equilibrium. Linearity is assumed so that the damping rate is not depending on the amplitude of the fluctuation. This simple picture is supported by classical molecular dynamics simulations, where starting from any initial state, after some relaxation time (even if we start with a homogeneous density, the pair correlation, e.g., must be established) the current-current correlation function is derived from the calculated trajectory in the $6\,N$ dimensional phase space ($\Gamma$ space). The transition from classical to quantum mechanics is done within the LRT based on the Zubarev NSO as presented in this work. 

Despite excellent results have been obtained comparing calculated conductivities with measured values for solids, liquids, and plasmas,
there are some open questions which demand a deeper understanding of the FDT. Some of them will be indicated here.

1) It is a miracle why linearity in the electrical field works, because the trajectories are weakly perturbed only for extremely weak fields,
and any realistic field leads to a entirely different trajectory (chaotic systems).
Nevertheless, the theory of irreversible processes assumes linear behavior also for realistic fields, and transport coefficients including the electrical conductivity are introduced this way. 
A possible answer is that the distribution function representing the ensemble is changing linearly with the (weak) 
field \cite{Zub69}, but then the question arises at which fields non-linearity may occur. 

2) Non-equilibrium is connected with irreversibility, at a definite time something happens (e.g. a broken glass) what is not possible 
considering a time-reversed movie. In particular, in LRT  the stationary case is of interest which is homogeneous in time.  
Also periodic fields and the corresponding frequency-dependent response are quasi-homogeneous in time. 
Because of linearity, an event occurring at a definite time (e.g. switch-off of an electrical field) can be decomposed by a Fourier transformation, and the response is also superposed from the corresponding Fourier components. 
The explicit instant of time where the event happens has no relevance (it is hidden in the phases of the Fourier component).
The distinction between past and future is not inherent in the time evolution of the system.

3) A serious problem is that irreversibility is connected with the production of entropy. 
This means that in the case of electrical conductivity  heat is produced. 
One can argue that this effect is of second order in the electrical field strength $E$ so it is not of relevance within LRT 
\cite{Vojta75}. However, this problem has to be considered.
The warming up is also seen in MD simulations when an external field is applied to the charged particle system.
In principle, we have to consider an open system coupled to a bath which absorbs the produced heat.
In the Zubarev NSO method considered here, it is the right hand of the extended von Neumann equation 
which contains the source term. We impose the stationary conditions so that $\rho_{\rm rel}$, in particular $T$, 
are not explicitly depending on time.
Then, the source term acts like an additional process describing the coupling to a bath without specifying the microscopic process.
The parameter $\epsilon$ has now the meaning of a relaxation time \cite{Zub70} and is no longer arbitrarily small but of the order $E^2$.

4) From a systematic microscopic point of view, one can introduce a process into the system Hamiltonian which describes the 
cooling of the system via the coupling to a bath, as known from the quantum master equations for open systems. 
Phonons related to the motion of ions can be absorbed by the bath, but one can calculate the electrical conductivity also for (infinitely) heavy ions so that the scattering of the electrons, accelerated by the field, is elastic. Collisions of electrons with the bath may help, 
but an interesting process to reduce the energy is radiation. Electrons which are accelerated during the collisions emit bremsstrahlung.
This heat transfers the gain of energy of electrons, which are moving in the external field, to the surroundings.

5) Can we really solve the problem this way, or is it only transferred to another object, the bath?
Has the source term a real significance \cite{Zub70}? Note that there are examples where  $\varepsilon$ is considered 
as finite to imitate a relaxation process, for instance to derive the Mermin result for the dielectric function \cite{Mermin}
to realize the particle number conservation or the hopping conductivity to realize the flow  of charged particles
across the border of the system \cite{ChrisRoep85}, see also \cite{DR}. However, any simulation of a real additional relaxation process not contained in $\Hop_{\rm S}$ should be considered as an approximation which has to be improved by a more fundamental description.\\

{\it Open systems: Coupling to the radiation field}.
The calculation of the conductivity of the Lorentz plasma model is a mystery because the 
Hamiltonian (\ref{ch6370}) is bilinear and can be diagonalized (in contrast to real
 collisions between charged particles). We have a scattering problem, and the use of the 
exact eigenstates will not explain irreversibility and the production of entropy.
Another mechanism is necessary to explain how a transport coefficient describing irreversible
behavior can be obtained.

A general approach to scattering theory was given by Gell-Mann and Goldberger \cite{GG} 
(see also Ref. \cite{Zubarev}) to incorporate the boundary condition into the 
Schr{\"o}dinger equation. The equation of motion in the potential ${\rm V}( \bs r)$ reads
\begin{equation}
\label{Gellmann}
 \frac{\partial}{\partial t}\psi_\epsilon({\bs r},t)+\frac{\ii}{\hbar}\Hop \,
\psi_\epsilon({\bs r},t)=- \epsilon[\psi_\epsilon({\bs r},t)-
\psi_{\rm rel}^{\hat t}({\bs r},t)].
\end{equation}
With $\Hop=\Hop_0+{\rm V}$, the relevant state is an eigenstate $|\bs p \rangle$ 
of $\Hop_0 $ which changes its value at the scattering time $\hat t$ where the 
asymptotic state $|\bs p' \rangle$ is formed. As known from the Langevin equation, 
one can consider $\psi_\epsilon({\bs r},t)=\varrho^{1/2} \exp (\ii S/\hbar)$ 
as a stochastic process \cite{r90,r2} related to a stochastic potential ${\rm V}( \bs r,t)$.
As an average, Eq. (\ref{Gellmann}) appears. The relaxation term is related to 
the fluctuations of ${\rm V}( \bs r,t)$. The average Hamiltonian dynamics 
is realized by the self-consistency conditions for $\psi_{\rm rel}^{\hat t}({\bs r},t)$,
see topic 2 of Sec. \ref{sec:NSO}.
The action ${\rm S}(\bs r,t)$ follows an equation of evolution
\begin{equation}
 \frac{\partial}{\partial t}{\rm S}(\bs r,t)+\frac{1}{2 m} (\nabla {\rm S})^2+{\rm V}( \bs r,t)
-\frac{\hbar^2}{2m} \frac{1}{\varrho^{1/2}}\varDelta \varrho^{1/2}=-\epsilon[{\rm S}(\bs r,t)
-S_{\rm rel}^{\hat t}(\bs r,t)]
\end{equation}
where $S_{\rm rel}^{\hat t}(\bs r,t)$ is the average action of the system eigenstates $\psi_{\rm rel}^{\hat t}({\bs r},t)$
formed at $\hat t$. 
In context with the extended von Neumann equation, it has been discussed also by Zubarev 
that the source term can be formulated for the exponent,
i.e. for $\ln \rho(t)$. Generalizing the Gell-Mann and Goldberger approach to implement the boundary conditions,
we can do it also for the action. 
As known from the Langevin equation, the relaxation term is connected with a stochastic process. 
The probability distribution $\varrho(\bs r,t)$ follows the equation of motion with the averaged action 
$S(\bs r,t)$
\begin{equation}
 \frac{\partial}{\partial t}{\varrho}(\bs r,t)+\frac{1}{ m}\nabla\cdot  (\varrho \nabla S)=0.
 \end{equation}

Of particular interest is the conductivity in the stationary case which is homogeneous in time. The system 
remains near thermodynamic equilibrium as long as the electrical field is weak so that the 
produced heat can be exported.
To describe a nonequilibrium state of a charged particle system (electrons and ions) with 
a stationary current (not superconducting),
we have to consider an open system. Mechanical work is imported, as described by the 
electrical field in the Hamiltonian, heat is exported. If the conductor is  embedded in 
vacuum, heat export is given by radiation. Bremsstrahlung is emitted during the collision of 
charged particles. Emission of photons can be considered as a measuring process to localize 
the charged particle during the collision process. The emitted power is
\begin{equation}
\frac{dH}{dt}=-\frac{2 e^2}{3 c^3} \left(\frac{d^2 {\bs r}}{dt^2}\right)^2+e \dot{\bs r} \cdot {\bs E}({\bs r},t).
\end{equation}
The emission and absorption of radiation, as described by a quantum master equation, is one of the possibilities to 
solve the problem of the export of entropy.
Such a master equation is connected with a stochastic process which describes the time evolution of the system,
for more discussion see Refs. \cite{r90,r2}.

	\section*{Acknowledgements}

The author thanks V. G. Morozov and N. M.  Plakida for valuable comments and discussions.

\appendix
\section{Linearization of the NSO}\label{App:1}
All  terms have to be evaluated in such a way, that
the total expression rests of order $\mathcal O(h)$. 
For the expression (\ref{2.205}), (\ref{2.206}) we  find after integration by parts
\bea
\label{apprho}
\rho_\epsilon(t)&=&\rho_{\rm rel}(t)-\int\limits_{-\infty}^t\dif{t_1}
\exx{\epsilon(t_1-t)}\op{U}(t,t_1)\left\{\frac{\ii}{\hbar} \left[(\Hop_{\rm S}+\Hop_{\rm F}^{t_1}),\rho_{\rm rel}(t_1)\right]
+\frac{\partial}{\partial t_1}\rho_{\rm rel}(t_1)\right\}\op{U}^\dagger(t,t_1).
\eea
Since $\Hop_{\rm S}$ commutes with $\rho_{\rm eq}$ (equilibrium!), the curly bracket is of order $\mathcal O(h)$.
In particular, we have for the first term the time derivative in the Heisenberg 
	picture,
	\begin{equation}
	\frac{\ii}{\hbar}[\Hop_{\rm S},\beta\int\limits_0^1\dif\lambda
	\sum\limits_nF_n(t_1)\op{B}_n(\ii\lambda\beta\hbar)\rho_{\rm eq}]= \beta\int\limits^1_0\mbox{d}\lambda
	\sum_nF_n(t_i)\dot{\op{B}}_n(\mbox{i}\lambda\beta\hbar)\rho_{\rm eq}.
	\end{equation}
	For the second term of the integral in Eq. (\ref{apprho}) we use Kubo's identity
	\be
	\label{003}
	\left[\op{B} ,\exx{\op{A} }\right]=\int\limits_0^1 \exx{\lambda \op{A} } \left[\op{B} ,\op{A}  \right] \exx{(1-\lambda)
	\op{A} } \mathrm{d}\lambda.
	\ee
so that
	\bea 
	&&\frac{\ii}{\hbar}[\Hop_{\it F}^{t_1},\rho_{\rm eq}]=-\beta e^{-\ii \omega t_1}\int\limits_0^1\dif\lambda\sum\limits_jh_j
	\dot{\op{A}}_j(\ii\lambda\beta\hbar)\rho_{\rm eq}.
	\eea
	The last term in the curly bracket can be rewritten as
	\begin{equation}
		\frac{\partial}{\partial t_1}\rho_{\rm rel}=\beta\int\limits_0^1\mbox{d}\lambda\sum_n\dot{F}_n(t_1)\op{B}_n(\mbox{i}\lambda\beta\hbar)\rho_{\rm eq}.
	\end{equation}
Because we restrict ourselves to the order $\mathcal O(h)$,  for the 
time evolution operator we have
$\op{U}(t,t_1)\simeq\exx{-\ii\op{H}_{\rm S}(t-t_1)/\hbar}.$

\section{Evaluation of the polarization function}
\label{PiGG}

Using the quasiparticle approximation, the spectral function has a Lorentzian form and the integrals can be performed 
(the real part of $\Sigma$ has been dropped):
\begin{eqnarray}
&&\sum_{z_\nu,z_\nu'}\Pi^{GG}({\bs p},\ii z_\nu,{\bs Q},\ii Z_\lambda,{\bs p}',\ii z_\nu')
=\sum_{z_\nu,z_\nu'}G({\bs p},\ii z_\nu)G({\bs p+Q},\ii z_\nu+\ii Z_\lambda)\delta_{z_\nu,z_\nu'}  \delta_{{\bs p},{\bs p}'}\nonumber \\
&&=\sum_{z_\nu}\int \frac{\dif \omega_1}{2 \pi} \int \frac{\dif \omega_2}{2 \pi}  
\frac{\hbar}{2 \tau_p}\frac{\hbar}{2 \tau_{\bs p+\bs Q}} 
\frac{1}{(\omega_1-E_{p})^2+1/\tau_p^2} \frac{1}{(\omega_2-E_{\bs p+\bs Q})^2+1/\tau_{\bs p+\bs Q}^2}
 \frac{1}{\ii z_\nu-\omega_1} \frac{1}{\ii  z_\nu-\ii Z_\lambda-\omega_2 }\delta_{{\bs p},{\bs p}'} \nonumber \\
 &&
 =\sum_{z_\nu}\frac{\pi^2\hbar^2/4}{(E_{p}-\ii  z_\nu+\ii /\tau_p)(E_{\bs p+\bs Q}+\ii Z_\lambda-\ii z_\nu+\ii /\tau_{\bs p+\bs Q})}\delta_{{\bs p},{\bs p}'}\,.
\end{eqnarray}
We perform the summation over $z_\nu$ so that (we repeat the integrals over the Lorentzian profiles)
\begin{eqnarray}
\label{Pigg}
&&\sum_{z_\nu,z_\nu'}\Pi^{GG}({\bs p},\ii z_\nu,{\bs Q},\ii Z_\lambda,{\bs p}',\ii z_\nu')
 = \frac{1}{4} \frac{f(E_{p})-f(E_{\bs p+\bs Q})}{-\ii Z_\lambda+E_{p}-E_{\bs p+\bs Q} +\ii /\tau_p+\ii /\tau_{\bs p+\bs Q}}.
\end{eqnarray}
The spectral density follows from the analytical continuation 
$\ii Z_\lambda \to \omega+\ii \epsilon$.

\section{Evaluation of the vertex contribution}
\label{app:vertex}

Considering $e-i$ collisions in Born approximation, the one-loop contribution to the effective interaction $\Gamma$ is
 \begin{eqnarray}
\Gamma({\bs p},\ii z_\nu,{\bs Q},\ii Z_\lambda,{\bs p}_1,\ii z_1)=V^2(|{\bs p}-{\bs p}_1|)\sum_{\bs k} 
\frac{f(E^{\rm ion}_{{\bs k}+{\bs p}_1})-f(E^{\rm ion}_{{\bs k}+{\bs p}})}{\ii z_\nu-\ii  z_1+E^{\rm ion}_{{\bs k}+{\bs p}_1}
-E^{\rm ion}_{{\bs k}+{\bs p}}}
 \end{eqnarray}
 In the following, the summation over $z_1$ is performed with the pole in $\Gamma({\bs p},iz_\nu,{\bs Q},iZ_\lambda,{\bs p}_1,i z_1)$ 
 because a Bose distribution without chemical potentials occurs,
 \begin{equation}
f(\ii z_\nu+E^{\rm ion}_{{\bs k}+{\bs p}_1}-E^{\rm ion}_{{\bs k}+{\bs p}})=n_B(E^{\rm ion}_{{\bs k}+{\bs p}_1}-E^{\rm ion}_{{\bs k}+{\bs p}}).
 \end{equation}
 All other poles will give higher orders in the density. We can transform $\left[f(E^{\rm ion}_{{\bs k}+{\bs p}_1})-f(E^{\rm ion}_{{\bs k}+{\bs p}})\right] 
 n_B(E^{\rm ion}_{{\bs k}+{\bs p}_1}-E^{\rm ion}_{{\bs k}+{\bs p}}) = f(E^{\rm ion}_{{\bs k}+{\bs p}_1})\left[1-f(E^{\rm ion}_{{\bs k}+{\bs p}})\right]$.
 In the remaining expression one has to replace $\ii z_1$ by $\ii z_\nu+E^{\rm ion}_{{\bs k}+{\bs p}_1}-E^{\rm ion}_{{\bs k}+{\bs p}}$. 
 In the adiabatic limit where the ions have a large mass, $E^{\rm ion}_{{\bs k}+{\bs p}_1}-E^{\rm ion}_{{\bs k}+{\bs p}}$ 
 becomes very small and will be neglected. This corresponds to elastic collisions of the electrons with ions. 
 The summation over ${\bs k}$ can be performed so that the ion number 
$N_{\rm ion}=n_{\rm ion} \Omega$ appears.
Now we have for the polarization function 
\begin{eqnarray}
\label{integraleq}
 && \Pi^{\rm BSE}({\bs p},\ii z_\nu,{\bs Q},\ii Z_\lambda,{\bs p}',\ii z_\nu')=
\Pi^{0}({\bs p},\ii z_\nu,{\bs Q},\ii Z_\lambda)
 \left\{\delta_{\bs p,\bs p'}\delta_{z_\nu,z_\nu'}+\sum_{\bs p_1}V^2(|{\bs p}-{\bs p}_1|) n_{\rm ion}
\Pi^{\rm BSE}({\bs p}_1,\ii z_\nu,{\bs Q},\ii Z_\lambda,{\bs p}',\ii z_\nu')\right\}, \nonumber\\
 && \Pi^{0}({\bs p},\ii z_\nu,{\bs Q},\ii Z_\lambda)
=\int \frac{\dif \omega_1}{2 \pi} \int \frac{\dif \omega_2}{2 \pi}  
 A({\bs p},\omega_1) A({\bs p-\bs Q},\omega_2)
 \frac{1}{\ii z_\nu-\omega_1} \frac{1}{\ii z_\nu-\ii Z_\lambda-\omega_2 } . 
\end{eqnarray}
The integral equation with respect to $z_1$ has been resolved because of the elastic scattering by the ions. The integral equation with 
respect to ${\bs p}_1$ has to be solved, and the calculation of the current-current correlation function has to be performed.


The quantity we are interested in is the current-current correlation function that is the average over ${\bs p}$ and ${\bs p}'$. We introduce a new function, the vertex function
\begin{equation}
{\bs F}({\bs p},\ii z_\nu,{\bs Q},\ii Z_\lambda)=\sum_{\bs p_1,\bs p',z_\nu'} {\bs p}' V^2(|{\bs p}-{\bs p}_1|)n_{\rm ion}
\Pi^{\rm BSE}({\bs p}_1,\ii z_\nu,{\bs Q},\ii Z_\lambda,{\bs p}',\ii z_\nu')\,.
\end{equation}

For ${\bs Q} \to 0$, in an isotropic system the only direction is by ${\bs p}$ so that 
\begin{equation}
{\bs F}({\bs p},\ii z_\nu,{\bs Q},\ii Z_\lambda)={\bs p} {\tilde F}(|{\bs p}|,\ii z_\nu,{\bs Q},\ii Z_\lambda).
\end{equation}
We use this to solve the integral equation for the vertex function,
\begin{eqnarray}
\label{pFtilde}
&& {\bs p} {\tilde F}(|{\bs p}|,\ii z_\nu,{\bs Q},\ii Z_\lambda) =\sum_{p_1} V^2(|{\bs p}-{\bs p}_1|)n_{\rm ion}
\int \frac{\dif \omega_1}{2 \pi} \int \frac{\dif \omega_2}{2 \pi}  
 A({\bs p}_1,\omega_1) A({\bs p_1-\bs Q},\omega_2)
 \frac{1}{\ii z_\nu-\omega_1} \frac{1}{\ii z_\nu-\ii Z_\lambda-\omega_2 } \nonumber\\ 
 && \times
 {\bs p}_1\left[1 +{\tilde F}(|{\bs p}|,\ii z_\nu,{\bs Q},\ii Z_\lambda)\right].
\end{eqnarray}
We made the assumption that ${\tilde F}(|{\bs p}_1|,\ii z_\nu,{\bs Q},\ii Z_\lambda)$ is a smooth function of $|{\bs p}_1|$. 
By reason of energy conservation the modulus of the momentum is not changed during scattering.
The solution reads
\begin{eqnarray}
\label{pFtilde1}
&& {\tilde F}(|{\bs p}|,\ii z_\nu,{\bs Q},\ii Z_\lambda) =-1 +\frac{1}{1-\frac{{1}}{p^2}
\sum_{\bs q} {\bs p}\cdot({\bs p+\bs q}) V^2(q)n_{\rm ion}\Pi^{0}({\bs p+\bs q},\ii z_\nu,{\bs Q},\ii Z_\lambda)}.
\end{eqnarray}

To calculate the dc conductivity (\ref{Kubo1}), we have ($Z_\lambda \to \omega'+\ii \epsilon$)
\begin{equation}
\label{Kubo2}
 \sigma^{\text{Kubo}}(\omega)= \lim_{Q \to 0}\frac{e^2 \beta}{3m^2 \Omega} \frac{\hbar}{\beta}\int \frac{\dif \omega'}{\ii \pi }\frac{1}{z-\omega'} \frac{1}{\omega'} \sum_{\bs p} p^2
{\rm Im}\,\left\{-\frac{1}{\beta}\sum_{z_\nu}\Pi^{0}({\bs p},\ii z_\nu,{\bs Q},\ii Z_\lambda)\left[1+{\tilde F}(|{\bs p}|,\ii z_\nu,{\bs Q},\ii Z_\lambda)\right]\right\}.
\end{equation}
We have to perform the summation over $z_\nu$. This can be done by iteration. We use here the approximation 
that the poles of $\Pi^{0}({\bs p},\ii z_\nu,{\bs Q},\ii Z_\lambda)$ are relevant which occur at $\ii z_\nu \to z  = E_{\bs p}, 
\ii Z_\lambda \to Z =E_{\bs p+{\bs q}}-E_{\bs p}$, using the quasiparticle approximation.
Then,
\begin{equation}
\label{Kubo3}
\Pi^{0}({\bs p},E_{\bs p},{\bs Q},E_{\bs p+\bs q}-E_{\bs p})=\int\frac{\dif \omega_1}{2 \pi}\int\frac{\dif \omega_1}{2 \pi}
\frac{1}{E_{\bs p}-\omega_1}\frac{1}{E_{\bs p+\bs Q}-\omega_2}
\frac{\hbar/\tau_p}{(\omega_1-E_{\bs p+\bs q})^2+(\hbar/2\tau_p)^2}
\frac{\hbar/\tau_p}{(\omega_2-E_{\bs p+\bs Q+\bs q})^2+(\hbar/2\tau_p)^2}.
\end{equation}
The propagators $1/(E-\omega)$ give vanishing principal values, and from the imaginary parts 
 $\delta$ functions occur which resolve the integrals.
Within the perturbation approach, we can replace one of the spectral functions by a quasiparticle spectral function.
The result is 
\begin{equation}
\label{Kubo4}
\Pi^{0}({\bs p},E_{\bs p},{\bs Q},E_{\bs p+\bs q}-E_{\bs p})=\pi \delta(E_{\bs p+\bs q}-E_{\bs p}) 4 \tau_p/\hbar\,.
\end{equation}
Insertion in (\ref{pFtilde1}) yields
\begin{eqnarray}
\label{pFtilde2}
&& {\tilde F}(|{\bs p}|,\ii z_\nu,{\bs Q},\ii Z_\lambda) =-1 +\frac{1}{1-
\sum_{\bs q} V^2(q)n_{\rm ion}\pi \delta(E_{\bs p+{\bs q}}-E_{ p}) 4 \tau_p/\hbar
+\sum_{\bs q}(q^2/2p^2) V^2(q)n_{\rm ion}\pi \delta(E_{\bs p+\bs q}-E_{\bs p}) 4 \tau_p/\hbar}.\nonumber\\&&{}
\end{eqnarray}
With $\sum_{\bs q} V^2(q)n_{\rm ion}2 \pi \delta(E_{\bs p+{\bs q}}-E_{\bs p})=\hbar/\tau_p$, Eq. (\ref{Imtau}), and
\begin{eqnarray}
\label{transp}
&& \sum_{\bs q}(q^2/2p^2) V^2(q)n_{\rm ion}\pi \delta(E_{\bs p+\bs q}-E_{\bs p})=\hbar/\tau^{\rm transp}_p
\end{eqnarray}
we arrive at 
\begin{eqnarray}
\label{pFtilde3}
&& {\tilde F}(|{\bs p}|,\ii z_\nu,{\bs Q},\ii Z_\lambda) =-1 +\frac{\tau^{\rm transp}_p}{\tau_p}.
\end{eqnarray}
Inserting in Eq. (\ref{Kubo2}) we find the result (\ref{Kubo5}).



\begin{thebibliography}{10}

\bibitem{Boltzmann}
L. Boltzmann, {\em Vorlesungen {\"u}ber Gastheorie} (Leipzig 1912), Bd.~2.

\bibitem{Bogoliubov}
N.N. Bogoliubov, {\em Problems of Dynamic Theory in Statistical Physics (in Russian)} (Gostekhizdat, Moscow-Leningrad 1946).

\bibitem{Zubarev}
D.N.  Zubarev, {\em Nonequilibrium Statistical Thermodynamics} (Plenum Press, New York, 1974);\\
Doklady Akademii Nauk SSSR {\bf 140}, 92 (1961).

\bibitem{ZMR1}
D. Zubarev, V. Morozov, and G. R\"opke, {\em Statistical Mechanics of
  Nonequilibrium Processes} (Akademie-Verlag, Berlin, 1996), Vol.~1.

\bibitem{ZMR2}
D. Zubarev, V. Morozov, and G. R\"opke, {\em Statistical Mechanics of
  Nonequilibrium Processes} (Akademie-Verlag, Berlin, 1997), Vol.~2.

\bibitem{ZubarevGF}
D.N.  Zubarev, 
Uspekhi Fizicheskikh Nauk (Soviet Physics Uspekh) {\bf 71}, 71 (1960).  i

\bibitem{Luzzi}
R. Luzzi {\it et al.}, {\it Statistical Irreversible Thermodynamics}, 
 Teor. Matem. Fiz., in press.

\bibitem{Kuzemsky}
A. L. Kuzemsky, {\it Metod neravnovesnaya statoperatora},  Teor. Matem. Fiz., in press.

\bibitem{Ryazanov}
V.V. Ryazanov, 
Eur. Phys. J. B {\bf 72}, 629 (2009).

\bibitem{Roep88}
G. R\"opke, Phys. Rev. A {\bf 38},  3001  (1988).

\bibitem{r90}
G. R\"opke,  {\em Nonequilibrium Statistical Mechanics} (in Russian), Mir, Moscow 1990.

\bibitem{Redmer97}
R. Redmer, Physics Reports {\bf 282},  36  (1997).

\bibitem{Reinholz05}
H. Reinholz, Annales de Physique (Paris), {\bf 30}, 1 (2005).

\bibitem{rr12}
H. Reinholz, and G. R\"opke, Phys. Rev. E {\bf 85},  036401 (2012).

\bibitem{Gocke}
C. Gocke and G. R\"opke, Theor. Math. Phys. {\bf 154}, 26 (2008).

\bibitem{Cheng}
C. Lin, C. Gocke, G. R\"opke, and H. Reinholz,
Phys. Rev. A {\bf 93}, 042711 (2016).

\bibitem{ChrisRoep85}
V. Christoph, and G. R\"opke, phys. stat. sol. (b) {\bf 131},  11  (1985).

\bibitem{rerrw00}
H. Reinholz, R. Redmer, G. R\"opke, and A. Wierling, Phys. Rev. E {\bf 62},
  5648  (2000).

\bibitem{Kubo66}
R. Kubo, J. Phys. Soc. Japan {\bf 12}, 570 (1957); Rep. Prog. Phys. {\bf 29},  255  (1966).

\bibitem{Roep98}
G. R\"opke, Phys. Rev. E {\bf 57},  4673  (1998).

 \bibitem{r2}
G. R\"opke, {\it{Nonequilibrium Statistical Physics}} (Wiley-VCH, Weinheim, 2013).

\bibitem{R81}
G. R\"opke, Theor. Math. Phys. {\bf 46}, 184 (1981).

\bibitem{Adams07}
J. Adams {\it et al.}, Phys. Plasmas  {\bf  14},  062303  (2007).

\bibitem{Mor99}
V. D. Morozov {\it et al.}, Ann. Phys. (N.Y.) {\bf 278}, 127 (1999).

\bibitem{ZM}
D. N. Zubarev, V. G. Morozov, I. P. Omelyan, and M. V. Tokarchuk,
   Theor. Math. Phys. {\bf 96}, 997 (1993).

\bibitem{MRSP}
V. G. Morozov and G. R\"opke, J. Stat. Phys. {\bf 102}, 285 (2001).

\bibitem{Roepke89}
G. R\"opke and R. Redmer, Phys. Rev. A {\bf 39},  907  (1989).

\bibitem{KB}
L. P. Kadanoff and G. Baym, {\it Quantum Statistical Mechanics}, (New York, W. A. Benjamin, 1962).

\bibitem{RD}
G. R\"opke and R. Der, 
 phys. stat. sol. (b) {\bf 92}, 501 (1979).

\bibitem{Spitzer53}
J. L.~Spitzer, and R. H\"arm, Phys. Rev {\bf 89},  977  (1953).

\bibitem{rrrr15}
H. Reinholz, G. R\"opke, S. Rosmej, and R. Redmer,
Phys. Rev. E {\bf 91}, 043105 (2015).

\bibitem{Kalashnikov}
V. P. Kalashnikov, Teor. Matem. Fiz. {\bf 34}, 412 (1978).

\bibitem{Mori}
H. Mori, Prog. Theor. Phys. {\bf 34}, 399 (1965).

\bibitem{Plakida}
A.A. Vladimirov, D. Ihle, and N. M. Plakida, Phys. Rev. B {\bf 85}, 224536 (2012).

\bibitem{Zub69}
I remember a discussion with D. N. Zubarev in 1969.

\bibitem{Vojta75}
Discussion with G. Vojta when applying the Zubarev NSO, around 1975.

\bibitem{Zub70}
This interpretation was always strictly refused by D.N. Zubarev 
in several discussions in 1969. He considered $\epsilon$ as a purely mathematical object to perform a definite limit and to select the retarded solution.

\bibitem{Mermin}
G. R\"opke, A. Selchow, A. Wierling, and H. Reinholz, Phys. Lett. A {\bf 260}, 365 (1999).

\bibitem{DR}
R. Der and G. R\"opke,
Phys. Lett. {\bf 95A}, 347 (1983).
%
\bibitem{GG}
M. Gell-Mann and M. L. Goldberger,
Phys. Rev. {\bf 91}, 398 (1953).




\end{thebibliography}
\end{document}